\documentclass[preprint,12pt]{elsarticle}

\usepackage{amssymb}
\usepackage{amsmath,bm}
\usepackage{tocloft}

\usepackage[colorlinks=true]{hyperref}  
\hypersetup{
    bookmarks=true,         
    unicode=false,          
    pdftoolbar=true,        
    pdfmenubar=true,        
    pdffitwindow=false,     
    pdfstartview={FitH},    
    pdfsubject={},   
    pdfcreator={},   
    pdfproducer={}, 
    pdfkeywords={} {} {}, 
    pdfnewwindow=true,      
    colorlinks=true,       
    linkcolor=magenta, 
    citecolor=blue,        
    filecolor=magenta,      
    urlcolor=blue           
} 

\journal{Physica C}
\usepackage{xcolor}

\begin{document}

\begin{frontmatter}



\title{The foot, the fan, and the  cuprate phase diagram:\\
Fermi-volume-changing quantum phase transitions}


\author{Subir Sachdev} 

\affiliation{organization={Department of Physics, Harvard University},
            city={Cambridge},
            postcode={02138}, 
            state={MA},
            country={USA}}

\begin{abstract}
A Fermi liquid with a ‘large’ Fermi surface (FL) can have a quantum phase transition to a spin density wave state (SDW) with reconstructed ‘small’ Fermi pockets. Both FL and SDW phases obey the Luttinger constraints on the volume enclosed by the Fermi surfaces. Critical spin fluctuations lead to spin-singlet $d$-wave pairing, as observed in the cuprates. Studies of the influence of spatial disorder on the FL-SDW quantum phase transition predict an extended quantum-critical Griffiths-type phase at low temperatures on the large Fermi surface side. These computations agree with the `foot' of strange metal transport, and recent low temperature neutron scattering observations on La$_{2-x}$Sr$_x$CuO$_4$.

However, this theory cannot explain the higher temperature pseudogap and the `fan' of strange metal behavior of the hole-doped cuprates. Here we need to consider underlying Fermi-volume-changing quantum phase transitions without symmetry breaking. Then the small Fermi surface phase does not obey the Luttinger constraint, and the pseudogap metal is described by thermal fluctuations above a `fractionalized Fermi liquid’ (FL*) or a `holon metal', with the descriptions related by a duality on a background spin liquid.
The quantum critical fan is described using a field theory for an underlying FL-FL* quantum phase transition in the presence of spatial disorder. This field theory can be
mapped to a form which can be analyzed using the methods of the Sachdev-Ye-Kitaev model. Such an analysis successfully models linear-in-temperature resistivity, optical conductivity and thermopower observations in the quantum critical fan.

The confinement crossover connecting these lower and higher temperature descriptions is also discussed.
\begin{center}
\href{https://arxiv.org/abs/2501.16417}{arXiv:2501.16417}; Physica C {\bf 633}, 1354707 (2025).
\end{center}

\end{abstract}







\end{frontmatter}

\tableofcontents



\section{Jan Zaanen and the cuprates}
\label{sec:intro}

\begin{figure}[t]
\centering
\includegraphics[width=4in]{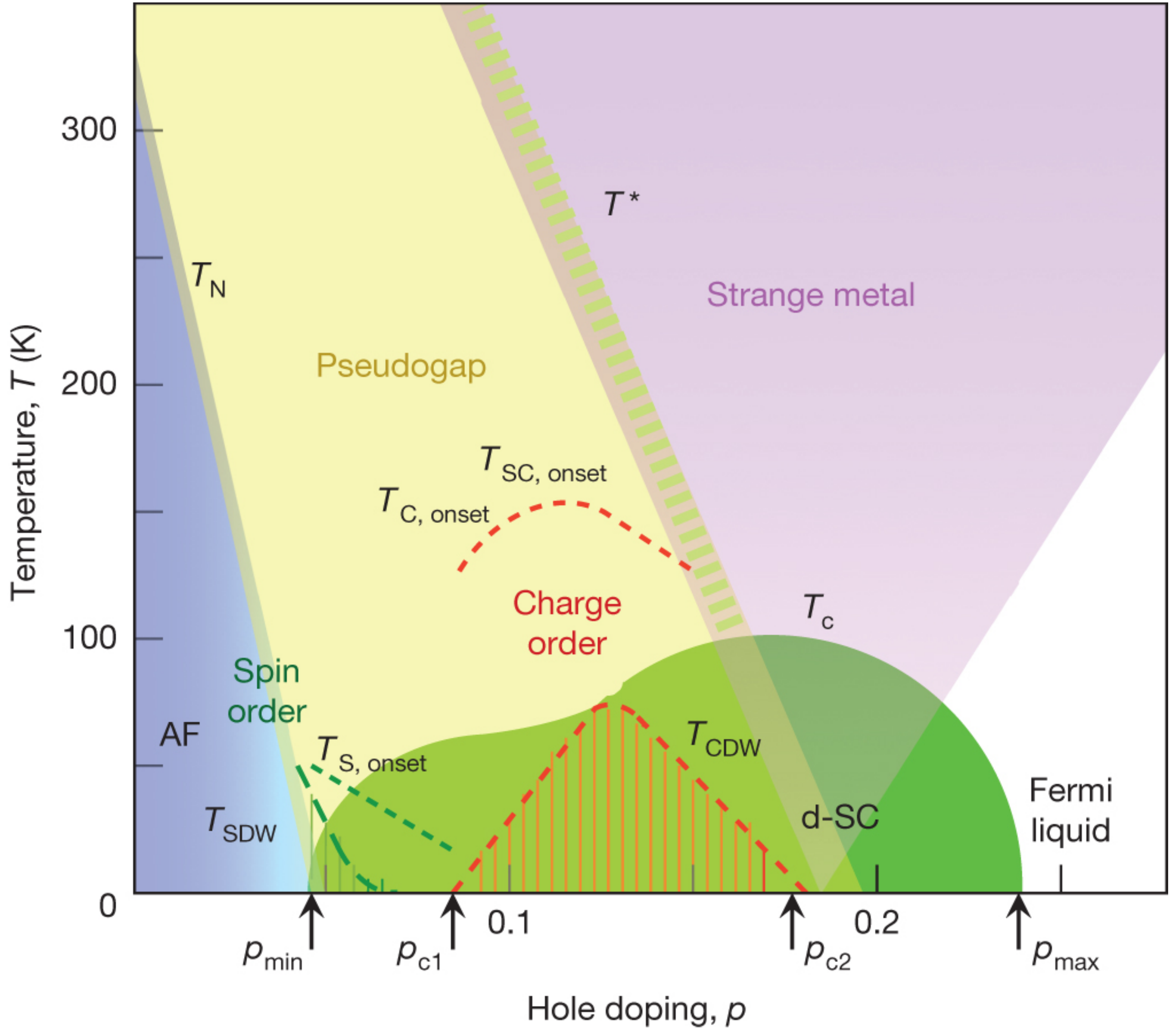}
\caption{Phase diagram of the hole-doped cuprate superconductors from Keimer {\it et al.\/}~\cite{phase_diag}.}
\label{fig:phase_diag}
\end{figure}

I met Jan Zaanen in the early nineties, and had innumerable discussions with him on a topic he was passionate about: the phase diagram of the cuprate high temperature superconductors. 
\begin{figure}
\centering
\includegraphics[width=3.5in]{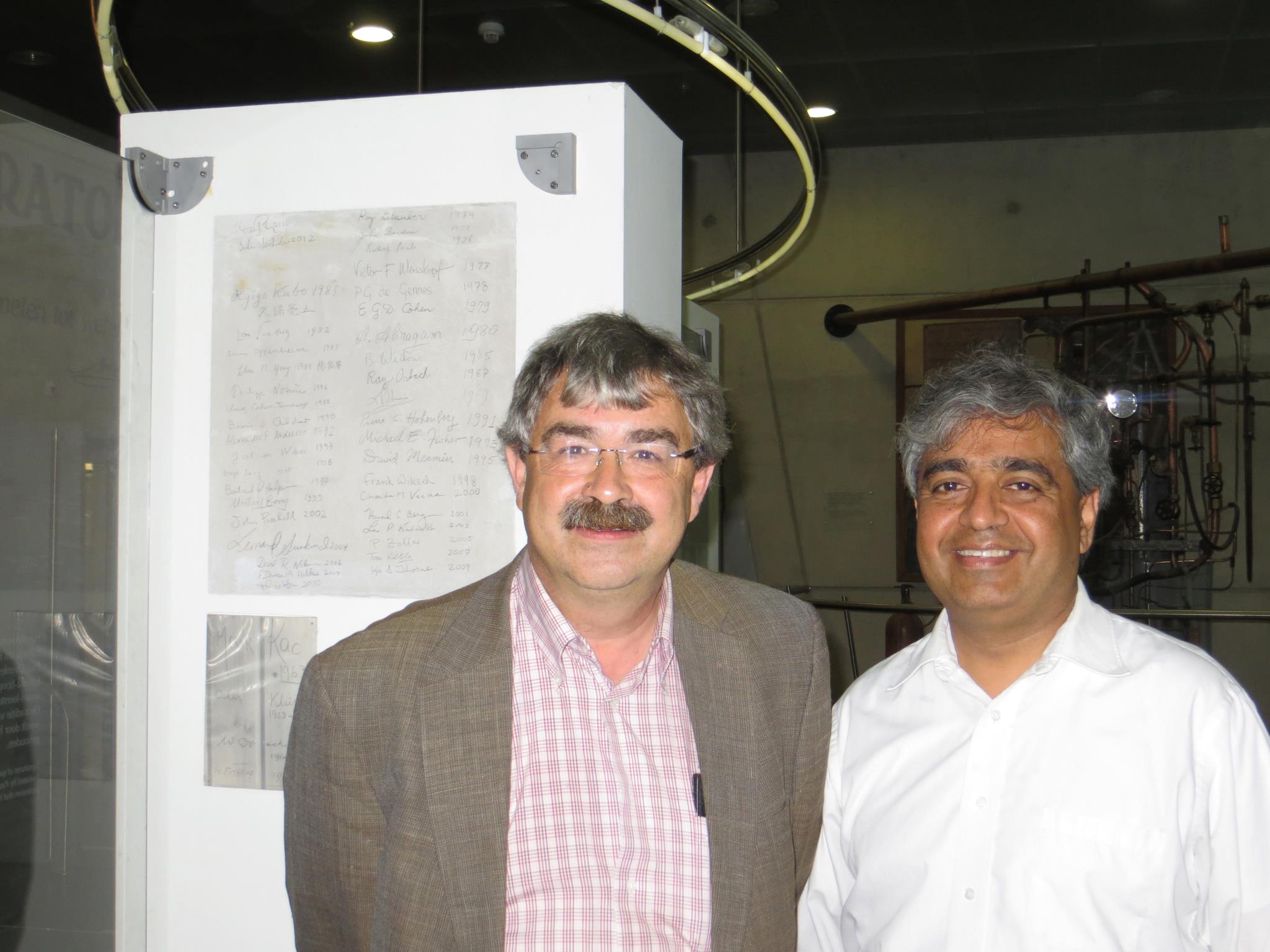}
\caption{Jan Zaanen and the author in 2012. In the background is the Lorentz wall of physicist signatures and the Kammerlingh Onnes refrigerator used for the discovery of superconductivity.}\label{fig:JZSS}
\end{figure}
Jan's pioneering early work on `stripe' charge density wave orders \cite{zaanen89} greatly influenced the subsequent experimental and theoretical developments, and eventually led to the phase diagram \cite{phase_diag} presented by Jan and collaborators shown in Fig.~\ref{fig:phase_diag}.
(See also the perspective of Proust and Taillefer \cite{Taillefer19}, which has also influenced the present article.)
Along with the observed low temperature phases with charge order and $d$-wave superconducting order, the diagram also shows two novel metallic phases, the pseudogap metal and the strange metal. Jan also thought a great deal about these metallic phases. He coined the term `Planckian' \cite{Zaanen04} to highlight the similarity between aspects of strange metals and charged black holes \cite{Herzog07,Hartnoll:2007ih}. 

From today's perspective, the important holographic model of strange metals by Jan and collaborators \cite{Zaanen09} can be presented as follows: the dopant electrons (visible in the `Fermi arc' spectrum of the pseudogap metal \cite{ShenShen05,Johnson11,Hoffman14,Davis14}) scatter strongly off a background two-dimensional black brane, leading to strange metal behavior. Specifically, they employed a charged black brane whose near-horizon physics is realized by an AdS$_2$ metric in 1+1 spacetime dimensions, with the spatial co-ordinate of AdS$_2$ being the emergent direction orthogonal to the black brane. Along with Refs.~\cite{Liu:2009dm,Faulkner:2009wj}, this was the first time I had heard of AdS$_2$ and its remarkable properties. Soon after, I realized \cite{Sachdev:2010um} that the properties of AdS$_2$ were strongly reminiscent of a class of models \cite{SY93} now known as Sachdev-Ye-Kitaev models. Specifically, the common features were \cite{Sachdev:2010um}:
\begin{itemize}
\item A 1+1 dimensional conformal structure of the Green's functions \cite{PG98,Schalm22} in a compressible system. This implied that a suitably defined relaxation time $\tau (\omega)$ obeyed Planckian scaling as a function of frequency, $\omega$, and temperature, $T$:
\begin{align}
\tau (\omega) = \frac{\hbar}{k_B T} F \left( \frac{\hbar \omega}{k_B T}\right) \,, \label{e1}
\end{align}
where $F$ is a scaling function which is precisely same for the SYK model and AdS$_2$, whose form is dictated by conformal invariance. A remarkable feature is that the relaxation time is independent of the underlying interaction strength between the electrons of the SYK model, in striking contrast to the Boltzmann equation of simple metals. 
\item
An extensive entropy in the limit of zero temperature \cite{GPS}. For a SYK model with $N$ sites, the precise statement for the entropy $S(T)$ is that
\begin{align}
\lim_{T \rightarrow 0} \lim_{N \rightarrow \infty} \frac{1}{N} S(T) = s_0\,, \label{e2}
\end{align}
where the order of limits is important and
\begin{align}
s_0 = \frac{\mathcal{G}}{\pi} + \frac{\ln 2}{4} = 0.464848\dots\,,
\end{align} 
with $\mathcal{G}$ Catalan's constant,
for a complex SYK model at half-filling.
Reissner-N\"ordstrom black holes with a total charge $\mathcal{Q}$ in 3+1 dimensions have a near horizon AdS$_2$ metric, and their entropy by the Hawking result
\begin{align}
S(T) = \frac{A (T) c^3}{4 \hbar G} \label{e3}
\end{align}
where $A(T)$ is the area of the horizon. The remarkable properties of black holes with a non-zero charge $\mathcal{Q}$ is that the area $A (T) \rightarrow A_0 \neq 0$ as $T \rightarrow 0$, with
\begin{align}
A_0 = \frac{2 G \mathcal{Q}^2}{c^4}\,. \label{e4}
\end{align}
Moreover connections of the SYK model entropy to that of the boundary entropy of the multi-channel Kondo problem \cite{ParcolletKondo} implied that the extensive zero temperature entropy of the SYK model was {\it not\/} realized by an exponential ground state entropy, but instead by an exponentially small spacing of energy levels down to an essentially non-degenerate ground state. Indeed the SYK model was the {\it first\/} to display this remarkable feature, which co-incides with the unusual thermodynamics of charged black holes. In contrast, all black holes that had been studied previously had a low energy supersymmetry, which led to an extensive degeneracy of the ground state \cite{Chowdhury:2021qpy}.
\end{itemize}
This connection between the SYK model and charged black holes has since seen extensive developments, and it has led to a deeper understanding of the quantum theory of black holes, including a precise expression for the low energy density of states of charged black holes in 3+1 dimensional Minkowski space \cite{Iliesiu:2022onk}; I have reviewed some of this elsewhere \cite{Sachdev:2023try}. These works mark the start of a period in which string theorists learnt from the physics of condensed matter models, reversing the earlier flow of information. 

Returning to applications of the SYK model to the cuprates, while the scaling result in (\ref{e1}) is precisely that observed in optical conductivity measurements in the cuprates \cite{Michon22} (see Fig.~\ref{fig:tau}; see also Refs.~\cite{Legros19,Ramshaw21}), 
the large entropies in (\ref{e2},\ref{e3},\ref{e4}) are far from any observations. Moreover, the AdS$_2$ holographic models (and also theories of critical Fermi surfaces of electrons coupled to gapless bosons \cite{Guo2022,Shi:2022toc}) have the artifact of a continuous translational symmetry, which leads to perfect electrical conductivity, rather than strange metal behavior \cite{Hartnoll:2007ih}. Much effort has been expended in overcoming these difficulties in applying these ideas to the cuprates, proceeding along two distinct directions. 

One direction modifies the holographic black brane models with either a periodic or random potential \cite{schalm2016,Hartnoll:2016apf,Zaanen23,Huang:2023ihu}. 

The other direction works directly with the SYK model, and modifies it to a more realistic representation of the underlying lattice scale physics, as reviewed in Ref.~\cite{Chowdhury:2021qpy}. It is this direction which we will discuss in the rest of this article. A fruitful formulation, which has been the focus of recent work, has been the mapping of theories of quantum phase transitions in disordered metals to the two-dimensional complex Yukawa-Sachdev-Ye-Kitaev (2dYSYK) model \cite{Fu16,Murugan:2017eto,Patel18,Patel:2018zpy,Marcus:2018tsr,Wang:2019bpd,Ilya1,Wang:2020dtj,Aldape22,WangMeng21,Esterlis21,Maria22,Patel:2022gdh,Guo2022,Schmalian1,Schmalian2,Alex23,Schmalian3,Guo24,PPS24,Serhii24,Li:2024kxr,Guo24a,Patel25,PLA24,XuHong24,LPS24,KimChatterjee25,SSORE}, which we will discuss in Sections~\ref{sec:qptsb} and \ref{sec:qptwsb}. 
With the benefit of hindsight, we can now see some of this progress could have been made after the early papers \cite{SY93,ParcolletKondo,PG98,GPS} on what was then sometimes called the SY model. But the intervening developments on the connections between SYK, black holes, and holography were important in pointing to the correct direction, and Jan played a significant role in these.
A possible holographic dual of the 2dYSYK model remains an interesting open question for future work, which we will not address here.

\section{The `foot' and the `fan'}
\label{sec:intro2}

\begin{figure}
\centering
\includegraphics[width=4in]{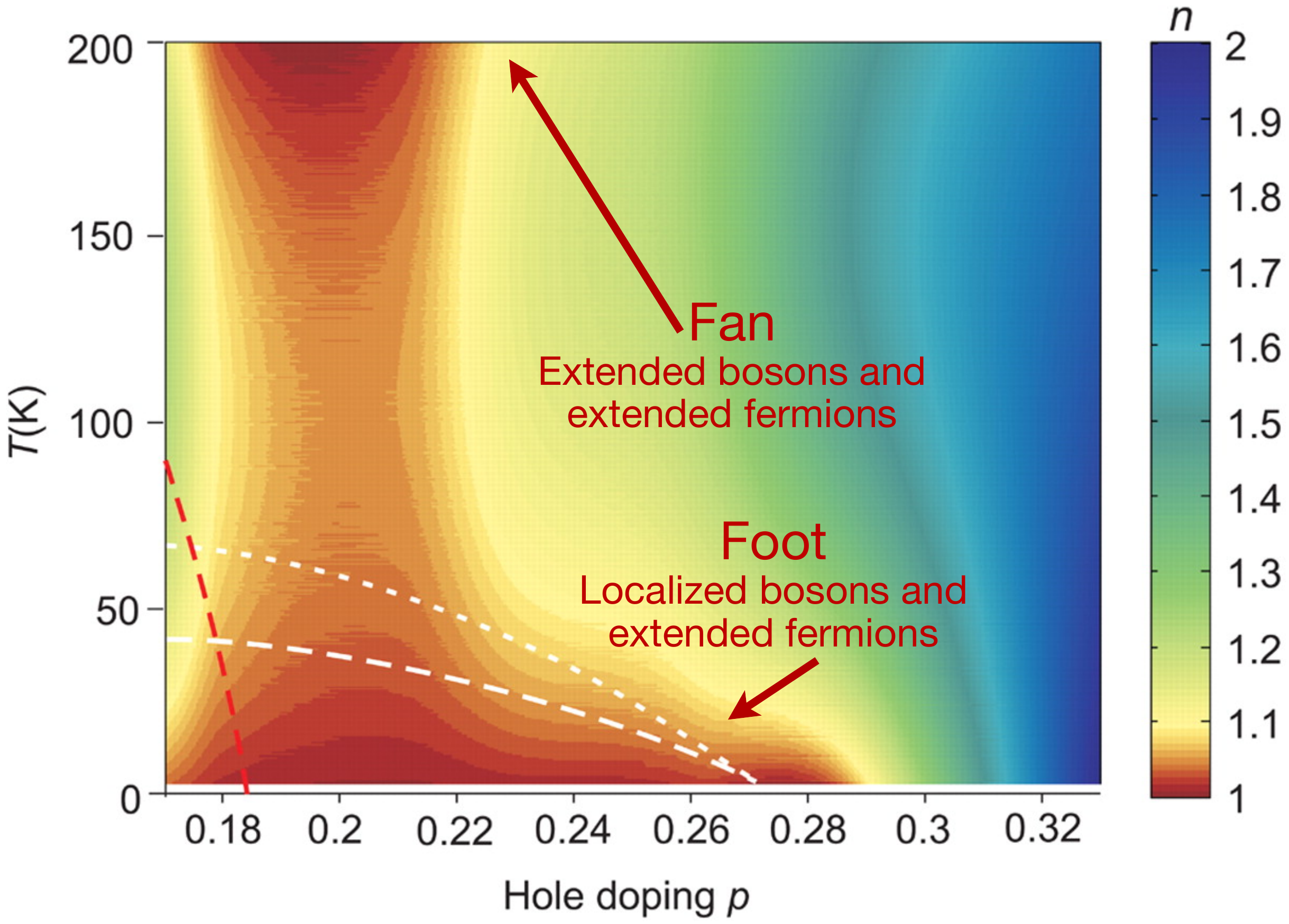}
\caption{Color density plot of the resistivity exponent $\rho (T) \sim T^n$ in La$_{2-x}$Sr$_x$CuO$_4$ from Cooper {\it et al.\/} \cite{Hussey_foot}. A magnetic field has been applied to suppress the superconductivity, and the resistance has been extrapolated to zero field. The `foot' and `fan' annotations have been added.}
\label{fig:footfan}
\end{figure}

Our discussion here of the phase diagram is motivated by Fig.~\ref{fig:phase_diag}, and the transport measurements of Cooper {\it et al.\/} \cite{Hussey_foot} in the hole-doped cuprates, shown in Fig.~\ref{fig:footfan} (there are similar observations also in the electron-doped cuprates \cite{Greene_rev}). 
Famously, the resistivity shows an extended `foot' of strange metal behavior at low $T$, along with a higher temperature quantum-critical `fan'. 
\begin{figure}[t]
\centering
\includegraphics[width=4.5in]{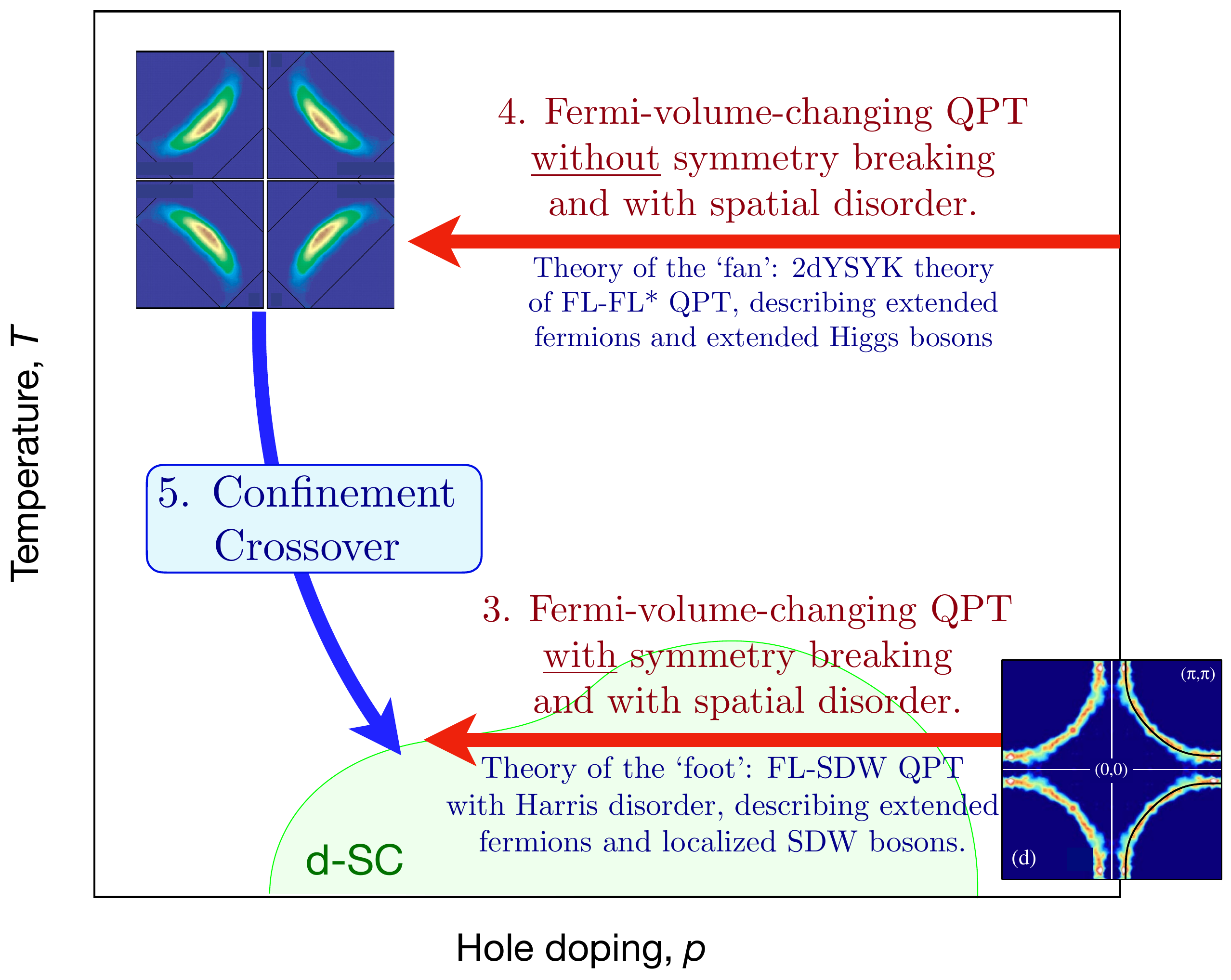}
\caption{Theories of the phase diagram of Fig.~\ref{fig:phase_diag}. The theories in Sections~\ref{sec:qptsb}, \ref{sec:qptwsb}, and \ref{sec:confinement} are labeled by the corresponding section numbers. The large Fermi surface photoemission plot is from Plat\'e {\it et al.}~\cite{Plate05}, and the pseudogap Fermi arcs are from Shen {\it et al.\/}~\cite{ShenShen05}. The latter is described by dual theories of thermal fluctuations of a holon metal or FL*.}
\label{fig:qpts}
\end{figure}
As displayed in Fig.~\ref{fig:qpts}, we explain these remarkable behaviors by quantum phase transitions (QPTs) involving Fermi volume change, one with symmetry breaking, and the other without. 

\begin{figure}[t]
\centering
\includegraphics[width=4in]{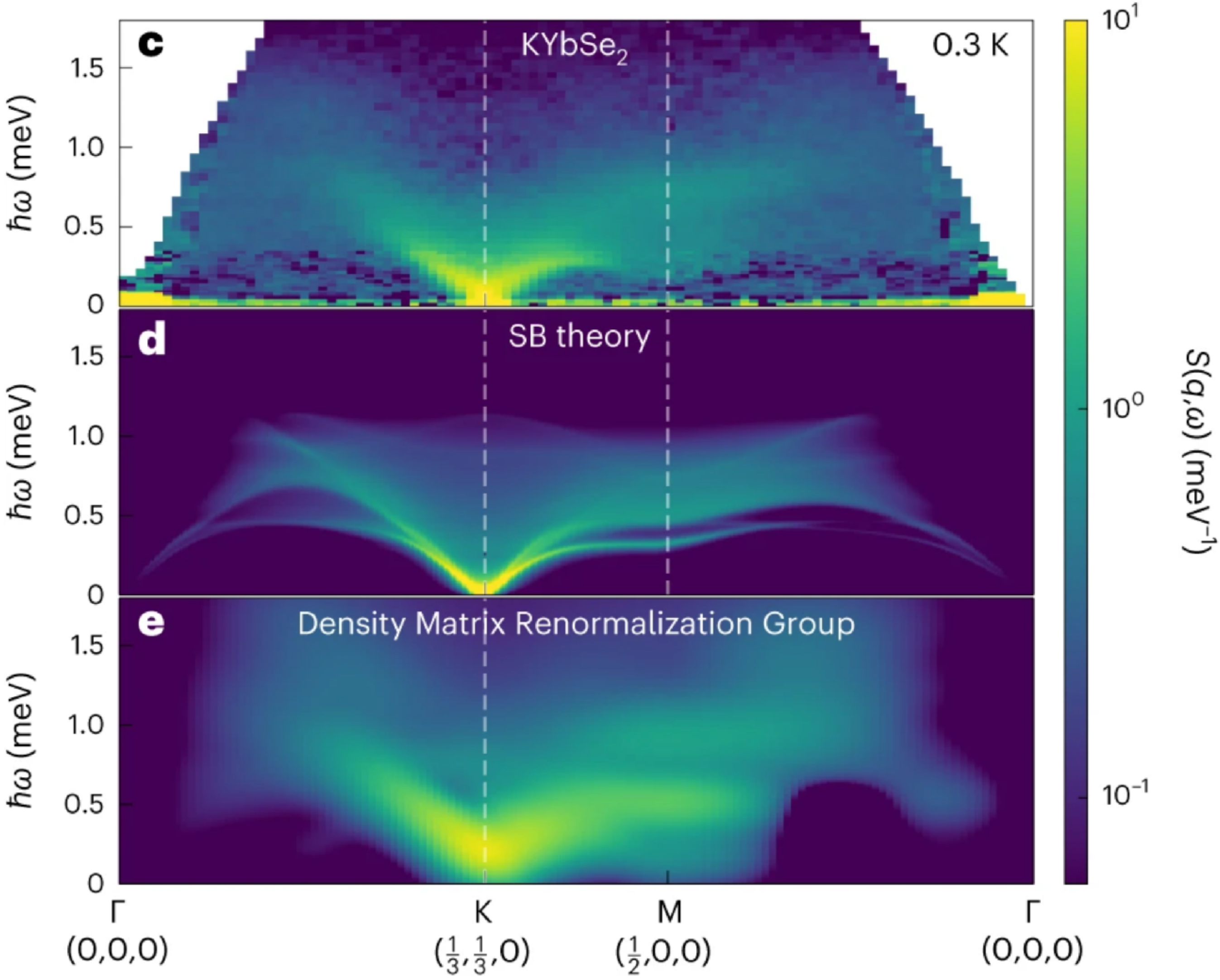}
\caption{From Scheie {\it et al.\/}~\cite{Scheie24}. Neutron scattering spectrum in the triangular lattice antiferromagnet KYbSe$_2$ (top panel), compared with the Schwinger boson (SB) theory of Ghioldi {\it et al.\/}~\cite{Batista18} and DMRG.}
\label{fig:KYbSe2}
\end{figure}

Before diving into the details of the cuprate phase diagram, it is helpful to consider simpler analogous phenomena which occur in the triangular lattice antiferromagnet KYbSe$_2$ \cite{Scheie24}. This compound has long-range magnetic order at low $T$, and associated spin waves are observed at low energies. However, at higher energies, the neutron scattering observations (see Fig.~\ref{fig:KYbSe2}) ``identify a diffuse continuum with a sharp lower bound within the measured spectra \ldots The key features of the data are reproduced by Schwinger boson theory'' \cite{Scheie24} of fractionalized spinons. 
(See also observations in NaYbSe$_2$ \cite{Scheie25}, 
which display a spinon continuum without long-range magnetic order, and are proposed to be a $\mathbb{Z}_2$ spin liquid \cite{NRSS91,Wen91}.)

We argue for a similar phenomenology for the cuprates in Fig.~\ref{fig:qpts}. At low temperatures and doping, we consider symmetry breaking with conventional spin and charge order.
At higher temperatures and low doping, we consider fractionalized states which enable Fermi volume change without symmetry breaking.

In Section~\ref{sec:qptsb} we discuss the low temperature `foot' in terms of a conventional spin density wave (SDW) onset in a disordered metal \cite{PPS24,PLA24}. The SDW order is closely connected to the stripe order discussed by Jan. 

However the SDW order disappears at higher temperatures, and so cannot explain the small Fermi surfaces (or `Fermi arcs') in the pseudogap regime. We describe this higher temperature pseudogap regime not in terms of one or more fluctuating conventional orders \cite{Kivelson_15,Lee_18}, but as a non-zero temperature realization of a quantum phase without symmetry breaking which could be stable at $T=0$ under suitable conditions. 
In principle, multiple quantum phases could exist, each distinct at  $T = 0$, yet all capable of providing satisfactory theoretical descriptions of the higher temperature pseudogap metal when thermal and quantum fluctuations are taken into account.
We consider theories of distinct $T=0$ states known as `fractionalized Fermi liquids' (FL*) \cite{TSSSMV03,TSSSMV04} and `holon metals' \cite{Lee89,ACL08,SDW09,Topo_PNAS,SS_ROPP19,Bonetti22} (see Fig.~\ref{fig:holon}). Both the $T=0$ FL* and holon metal states have Fermi surfaces which do not enclose the Luttinger volume, and such behavior is only possible in the presence of a background spin liquid \cite{TSSSMV03,TSSSMV04,APAV04,Bonderson16}. The 
spin liquid takes dual forms in the two phases, as we will discuss in Section~\ref{sec:confinement}. We present an underlying FL-FL* QPT in Section~\ref{sec:qptwsb}, and argue that it provides a satisfactory description over the higher temperature quantum critical fan \cite{LPS24}. But we do not fully resolve the nature of the pseudogap metal between FL* and the holon metal.

Additionally, as we lower the temperature from the pseudogap, there is a crossover to the conventional SDW or $d$-wave superconducting (d-SC) states at low $T$ (which are not described by the FL-FL* QPT theory): this requires confinement of the fractionalized excitations of the spin liquid background of the pseudogap, and is discussed in Section~\ref{sec:confinement}.

\section{The `foot': QPT with symmetry breaking}
\label{sec:qptsb}

For the QPT with symmetry breaking in Fig.~\ref{fig:qpts}, we consider the onset of conventional SDW order in a FL, as discussed originally by Hertz \cite{Hertz1976} and Millis \cite{Millis1993}, but in the presence of spatial randomness.

We write the spin density of the stripe order as ($a = x,y,z$)
\begin{align}
S_a ({\bm r}) = \sum_\ell \phi_{\ell a} e^{i {\bm K}_\ell \cdot {\bm r}}
\end{align}
where $\ell = 1 \dots 4$ labels the 4 ordering wavevectors ${\bm K}_\ell$ at $(\pi (1 \pm \delta), \pi)$ and $(\pi, \pi(1 \pm \delta))$. We are interested in fluctuations of the SDW order parameters $\phi_{\ell a}$ coupled to electrons $c_{{\bm k} \sigma}$ with dispersion $\varepsilon ({\bm k})$ which has a Fermi surface. 
We describe this with a 2dYSYK model \cite{Esterlis21,Patel:2022gdh} with imaginary time ($\tau$) Lagrangian
\begin{align}
\mathcal{L}_1 & = \sum_{{\bm k}} c_{{\bm k}\sigma}^\dagger \left( \frac{\partial}{\partial \tau} + \varepsilon ({\bm k})\right) c_{{\bm k}\sigma}  +
\int d^2 {\bm r} \, \Bigl\{
s \, [\phi({\bm r})]^2  \nonumber \\
& +  \sum_{\ell} [g + {\color{purple} g' ({\bm r})}] c^{\dagger}_\sigma ({\bm r})  \tau^a_{\sigma \sigma'} c_{\sigma'} ({\bm r}) \, \phi_{\ell a} ({\bm r}) e^{i {\bm K}_\ell \cdot {\bm r}} + K \, [ \nabla_{\bm r} \phi ({\bm r})]^2 +u  \,[\phi({\bm r})]^4 \nonumber \\
& ~~~~~~~~~+ {\color{purple} v({\bm r})} c_\sigma^\dagger ({\bm r}) c_\sigma  ({\bm r}) \Bigr\} \,. \label{e11}
\end{align}
Here $\tau^a$ are the Pauli matrices, $s$ is the parameter employed to tune across the transition, and $g$ is the Yukawa coupling between the fermions and bosons. We have included two sources of spatial randomness (symbols in purple are fixed random functions of space, with no dynamics). The
spatially random potential {\color{purple} $v ({\bm r})$}, with ensemble averages {\color{purple} $\overline{v({\bm r})} = 0$, $\overline{v({\bm r}) v({\bm r'})} = v^2 \delta({\bm r}-{\bm r'})$}, acts on the fermion density, and plays a central role in the theory of disorder-induced electron localization \cite{TVRRMP}. Such fermion localization effects are also present here, but all indications are that such effects are not important for the cuprates. 

Instead, our focus will be on the more relevant `Harris disorder', induced by spatial randomness in the position of the quantum critical point. This is represented in (\ref{e11}) by the 
spatially disordered Yukawa coupling {\color{purple} $g' ({\bm r})$} with {\color{purple} $\overline{g' ({\bm r})} = 0$, $\overline{g' ({\bm r}) g' ({\bm r'})} = g'^2  \delta({\bm r}-{\bm r'})$}. A more conventional form of the Harris disorder is in terms of a `random mass' under which the coupling $s \rightarrow s +  {\color{purple} \delta s ({\bm r})}$; but this can be mapped to {\color{purple} $g' ({\bm r})$} by a rescaling of $\phi$ chosen to make the co-efficient of $[\phi ({\bm r})]^2$ independent of ${\bm r}$. In either form, the Harris disorder leads to localization of the boson $\phi$ at low energies (as we discuss below), and so must be treated non-perturbatively. However, in the higher energy regime where the bosons are extended, it is preferable to place the Harris disorder only in {\color{purple} $g' ({\bm r})$} as this has the salutary effect of accounting for the spatial structure of the bosonic eigenmodes, and so enables use of the self-consistent SYK methods for the spatially averaged Green's functions.

Patel {\it et al.\/} \cite{PLA24} have recently studied the 2dYSYK model (\ref{e11}) at $g=0$ by large scale, high precision quantum Monte Carlo simulations (see also Ref.~\cite{Patel25}). Here we discuss the analysis of Ref.~\cite{PPS24} which yields similar results: it applies a Hartree-Fock approximation to the $\phi^4$ term in an Hertz effective theory for the bosons alone, but treats disorder numerically exactly.
We integrate out the fermions from (\ref{e11}) (assuming fermionic eigenmodes remain extended), and consider the Landau-damped Hertz theory for the boson alone.
The disorder in {\color{purple}$g' ({\bm r})$} will lead to spatial disorder in all couplings in the Hertz theory, but we retain only the most relevant `random mass' disorder {\color{purple} $\delta s ({\bm r})$}. We discretize on a lattice of sites (labeled by $j$), and write the SDW order parameters in terms of a real $\phi$ with a single index $a = 1 \ldots M$ with $M=12$. In this manner, we obtain the action
\begin{align}
\mathcal{S} = & \mathcal{S}_\phi +  \mathcal{S}_{\phi d} \nonumber \\
    \mathcal{S}_\phi =& \int d \tau \Biggl[ \frac{J}{2}\sum_{\langle i j \rangle} \left( \phi_{i a} -\phi_{ja} \right)^2 +  \sum_{j} \biggl\{ \frac{s+ {\color{purple} \delta s_j}}{2} \phi_{j a}^2   + \frac{u}{4M} \left( \phi_{j a}^2 \right)^2 \biggr\}\Biggr] \nonumber \\
    \mathcal{S}_{\phi d} =& \frac{T}{2} \sum_{\Omega} \sum_j \left(\gamma |\Omega| + \Omega^2/c^2 \right)|\phi_{ja} (i\Omega)|^2 \,, \label{e10}
\end{align}
where $\Omega$ is a Matsubara frequency at a temperature $T$, $\gamma$ is the Landau damping, and the $\Omega^2/c^2$ term has been inserted as a high frequency cutoff. For simplicity, we have assumed a global O($M$) symmetry, but this assumption can relaxed without significantly modifying the results.

The theory in (\ref{e10}) has been studied using a strong disorder renormalization group \cite{TV07,TV09,TV13}. But the same basic results are obtained by the method of Ref.~\cite{PPS24} (originally used for a related problem in $d=1$ in Ref.~\cite{Adrian08}), which also allows study of the crossover at higher energies to weak disorder, and this will be important for our purposes. Following Refs.~\cite{Adrian08,PPS24}, we replace $\mathcal{S}_\phi$ by an effective quadratic action, while renormalizing the space dependent mass in a self-consistent manner; this leads to 
\begin{align}
    \widetilde{\mathcal{S}}_\phi =& \int d \tau \Bigl[ \frac{J}{2}\sum_{\langle i j \rangle} \left( \phi_{i a} -\phi_{ja} \right)^2 +  \sum_{j} \frac{ {\color{purple} \widetilde{s}_j}}{2} \phi_{j a}^2   \Bigr] \nonumber \\
    & {\color{purple} \widetilde{s}_j} = s + {\color{purple} \delta s_j} + \frac{u}{M}
    \sum_a \left\langle \phi_{ja}^2 \right\rangle_{\widetilde{\mathcal{S}}_\phi + \mathcal{S}_{\phi d}} \nonumber \\
   &~~~ = s + {\color{purple} \delta s_j} + u T \sum_\Omega \sum_{\alpha} \frac{\psi_{\alpha i} \psi_{\alpha j}}{\gamma |\Omega| + \Omega^2/c^2 + e_\alpha}\,, \label{e12}
\end{align}
where $e_\alpha$ and $\psi_{\alpha j}$ are eigenvalues and eigenfunctions of the $\phi$ quadratic form in $\widetilde{\mathcal{S}}_\phi$, labeled by the index $\alpha = 1 \ldots L^2$ for a $L \times L$ sample. For each disorder realization {\color{purple} $\delta s_j$}, the values of {$\color{purple} \widetilde{s}_j$} are determined by numerically solving (\ref{e12}), and this also yields results for the eigenvalues $e_\alpha$ and the eigenvectors $\psi_{\alpha j}$. 
\begin{figure}[t]
\centering
\includegraphics[width=3.5in]{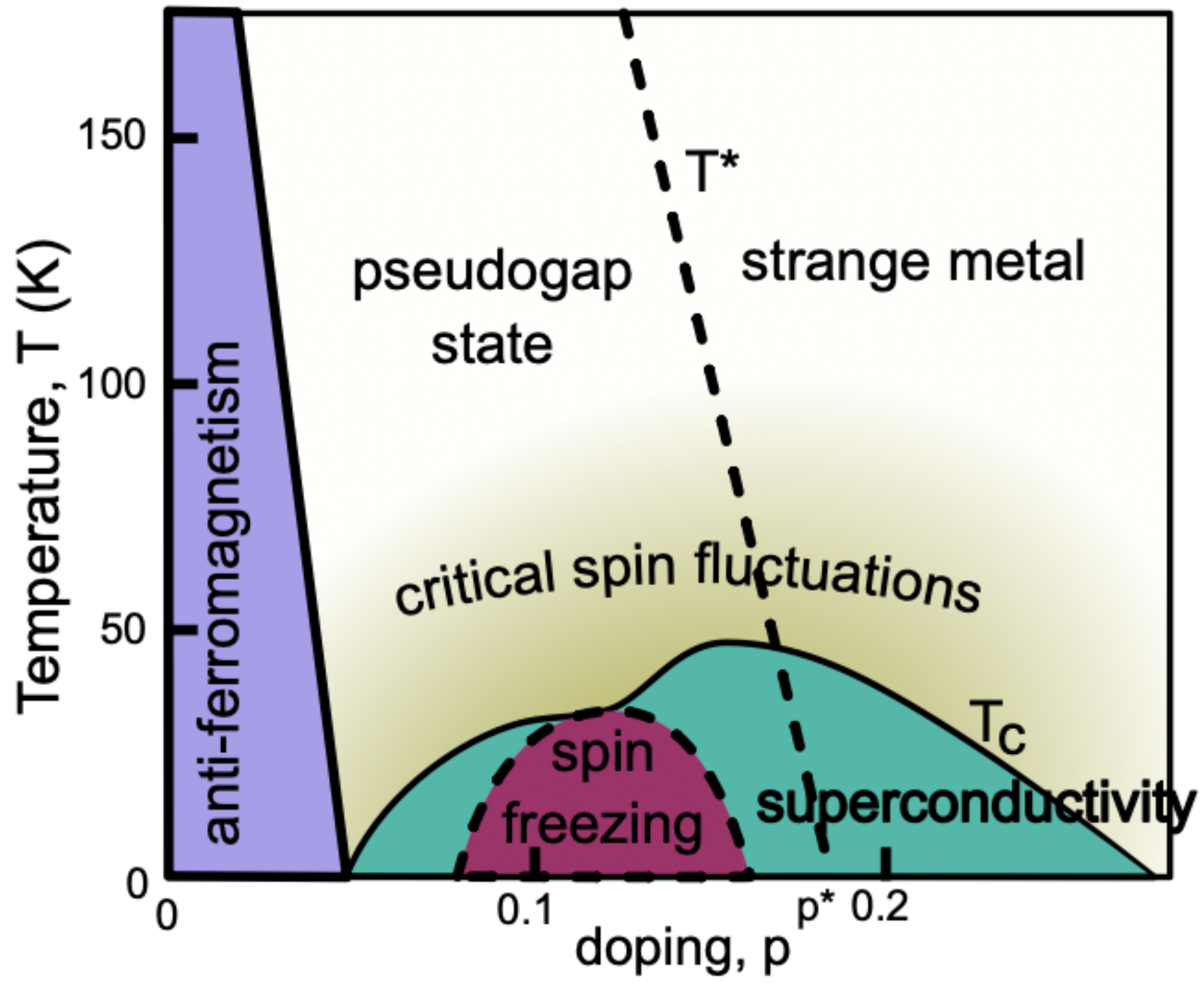}
\caption{Phase diagram of the hole-doped cuprates from Radaelli {\it et al.\/}~\cite{Hayden25}, based on their neutron scattering observations.}
\label{fig:Hayden}
\end{figure}

The results \cite{TV07,TV09,TV13,Adrian08,PPS24,PLA24} show a $T=0$ FL-SDW QPT at some $s=s_c$, accompanied by an extended gapless quantum Griffiths phase for $s > s_c$, with physics similar to the random Ising model in a transverse field \cite{Motrunich00}. Fermion spectral functions and transport properties have been computed, and are qualitatively similar to the `foot' of the strange metal in Fig.~\ref{fig:footfan}.
We note that a similar localized boson foot regime has also been found in the non-pertubative quantum Monte Carlo at $g=0$ \cite{PLA24}. 

These results are consistent with the recent high field observations of Campbell {\it et al.\/} \cite{Proust24} relating strange metals to spin fluctuations. Tranquada {\it et al.\/} \cite{Tranquada24} have also connected strange metal behavior to inhomogeneous spin fluctuations. 
Radaelli {\it et al.} \cite{Hayden25} have recently observed an extended regime of critical spin fluctuations in La$_{2-x}$Sr$_x$CuO$_4$ (extending earlier observations \cite{Aeppli93,Aeppli98,Hayden23}; see Fig.~\ref{fig:Hayden}), and their results are in general agreement with analytic continuation in frequency of the results of Refs.~\cite{PPS24,PLA24}. Similar neutron scattering results appeared recently in Ref.~\cite{Chang25}.

These neutron scattering results are likely closely related to EELS observations of singular density fluctuations \cite{Mitrano18,Husain19,Abbamonte24,Philip24}. 
There have been computations of density fluctuations in spin fluctuation models \cite{Joshi20, Chowdhury23} related to (\ref{e10}), but they have not yet included contributions of localized modes.

It would be interesting to study the origin of $d$-wave superconductivity from these disordered spin fluctuations, extending the theories of $d$-wave superconductivity from a clean Fermi liquid \cite{scalapino1986d,Raghu10}.

We conclude this discussion of the FL-SDW QPT in a disordered metal by highlighting the structure of the bosonic eigenmodes $\psi_\alpha$ near the quantum critical point.
\begin{figure}
\begin{center}
\includegraphics[width=4.5in]{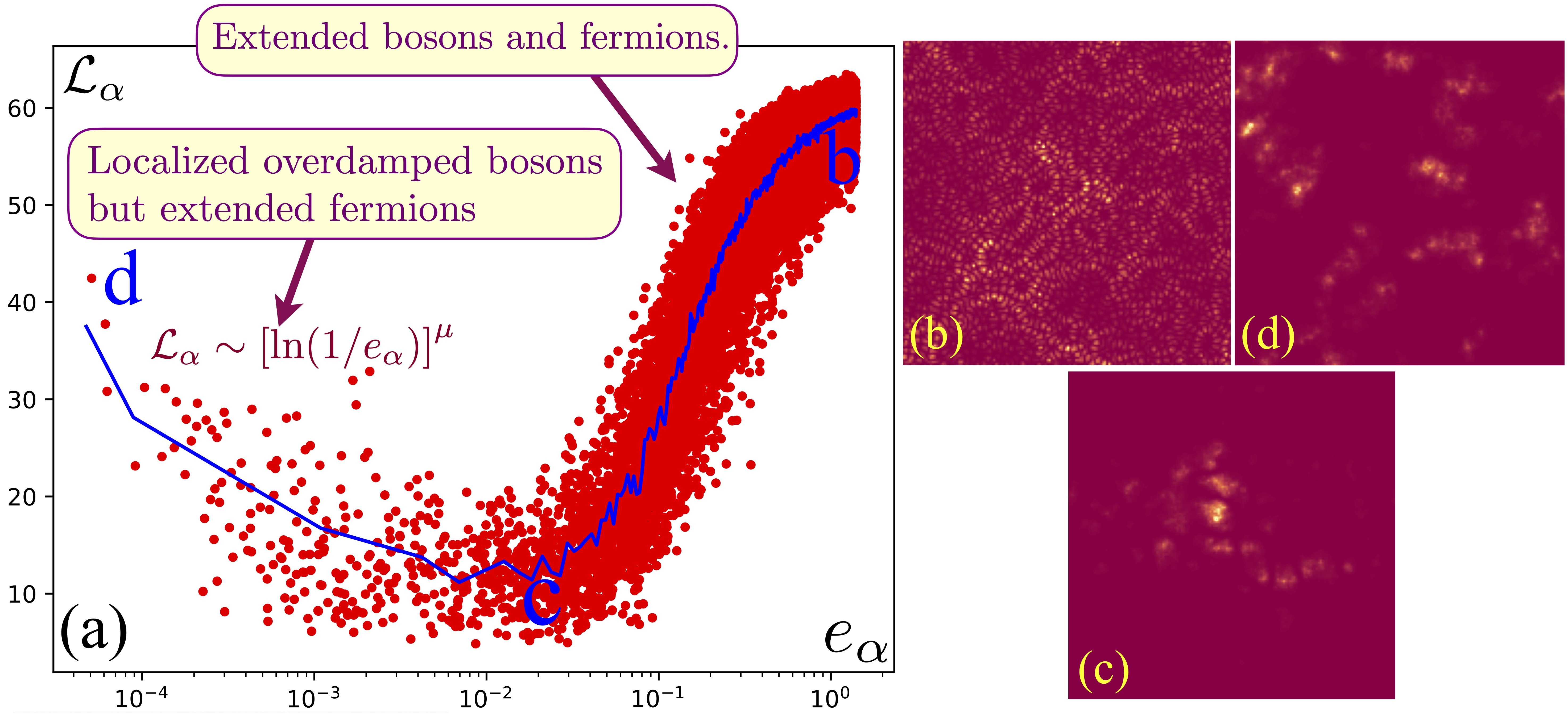}
\end{center}
\caption{Adapted from Patel {\it et al.\/}~\cite{PPS24}. (a) Localization length $\mathcal{L}_\alpha$ of overdamped bosonic eigenmodes of $\widetilde{\mathcal{S}}_\phi$ in (\ref{e12}) as a function of their energy $e_\alpha$. (b,c,d) Pictures of the corresponding bosonic eignfunctions.  The `foot' is described by the localized bosons.
The universal theory of the `fan' by Patel {\it et al.\/}~\cite{Patel:2022gdh} and Li {\it et al.\/}~\cite{Li:2024kxr} employs the mapping to the 2dYSYK model in (\ref{e22}), and applies to the regime of extended bosons.}
\label{fig:eigenmodes}
\end{figure}
Typical results from Ref.~\cite{PPS24} are shown in Fig.~\ref{fig:eigenmodes}. At higher $e_\alpha$, we obtain extended bosonic eigenmodes, where disorder can be treated in a self-averaging manner, as will be discussed near (\ref{SYKeq}). However, we observe a remarkable non-monotonic behavior in the localization length $\mathcal{L}_\alpha$ of the eigenmodes as $e_\alpha$ is reduced, which requires a non-perturbative and non-self-averaging treatment of disorder in either the random mass form of (\ref{e11}), or the random coupling form of (\ref{e22}).
The bosonic modes localize at an intermediate energy, but then the localization length increases logarithmically upon further lowering the energy. This logarithmic increase is characteristic of infinite-randomness fixed points \cite{Motrunich00,TV13}, and arises from the interaction of the incipient localized modes. It is these localized modes of SDW spin fluctuations in Fig.~\ref{fig:eigenmodes} which are responsible for the extended critical quantum Griffiths phase, and they lead to the `foot' in Fig.~\ref{fig:footfan}, and the critical spin fluctuations of Fig.~\ref{fig:Hayden} \cite{PPS24,PLA24}. 

We emphasize that this localization of bosonic modes takes place while the fermionic eigenstates remain extended. This is similar to early work on the disordered Hubbard model in three dimensions \cite{Milica89}. The localized SDW modes also have some resemblance to models of the strange metal using two-level systems \cite{Berg24a,Berg24b,Bashan25}.

It would clearly be of interest in future experiments to obtain spatially resolved spin fluctuation spectra, and then determine if the critical spin fluctuations observed in neutron scattering \cite{Hayden25} have the spatial structure illustrated in Fig.~\ref{fig:eigenmodes}. Inhomogeneity in the superconductivity has already been observed in scanning tunneling microscopy experiments \cite{Davis08,Allan23} in work with Jan playing a significant role.

\section{The `fan': QPT without symmetry breaking}
\label{sec:qptwsb}

In principle, the symmetry breaking theory for the `foot' in (\ref{e11}) can be extended to higher temperatures to also provide a theory for the `fan', and this may well be the appropriate approach for many correlated electron compounds \cite{Tremblay04}. The theory of the `fan' in Refs.~\cite{Patel:2022gdh,Li:2024kxr} using the disorder averaged in (\ref{SYKeq}) below also applies to the extended SDW bosons in Fig.~\ref{fig:eigenmodes}.

However, such an approach does not apply to the hole-doped cuprates in particular, because the spin correlation length in the pseudogap regime is too small to explain the large gap seen in photoemission in the anti-nodal region of the Brillouin zone \cite{Chubukov23,Topo_PNAS}. 
Therefore, as discussed in Section~\ref{sec:intro2}, we model the pseudogap metal of the hole-doped cuprates by a metallic state in which the electrons missing from the Fermi surface count are placed in a spin liquid with fractionalization \cite{TSSSMV03,TSSSMV04,APAV04,Bonderson16}. This leads to a FL-FL* QPT description of the higher temperature quantum critical fan, which we discuss in the present section. Further discussion of the nature of the pseudogap metal itself is deferred to Section~\ref{sec:confinement}.
\begin{figure}
\begin{center}
\includegraphics[width=3.9in]{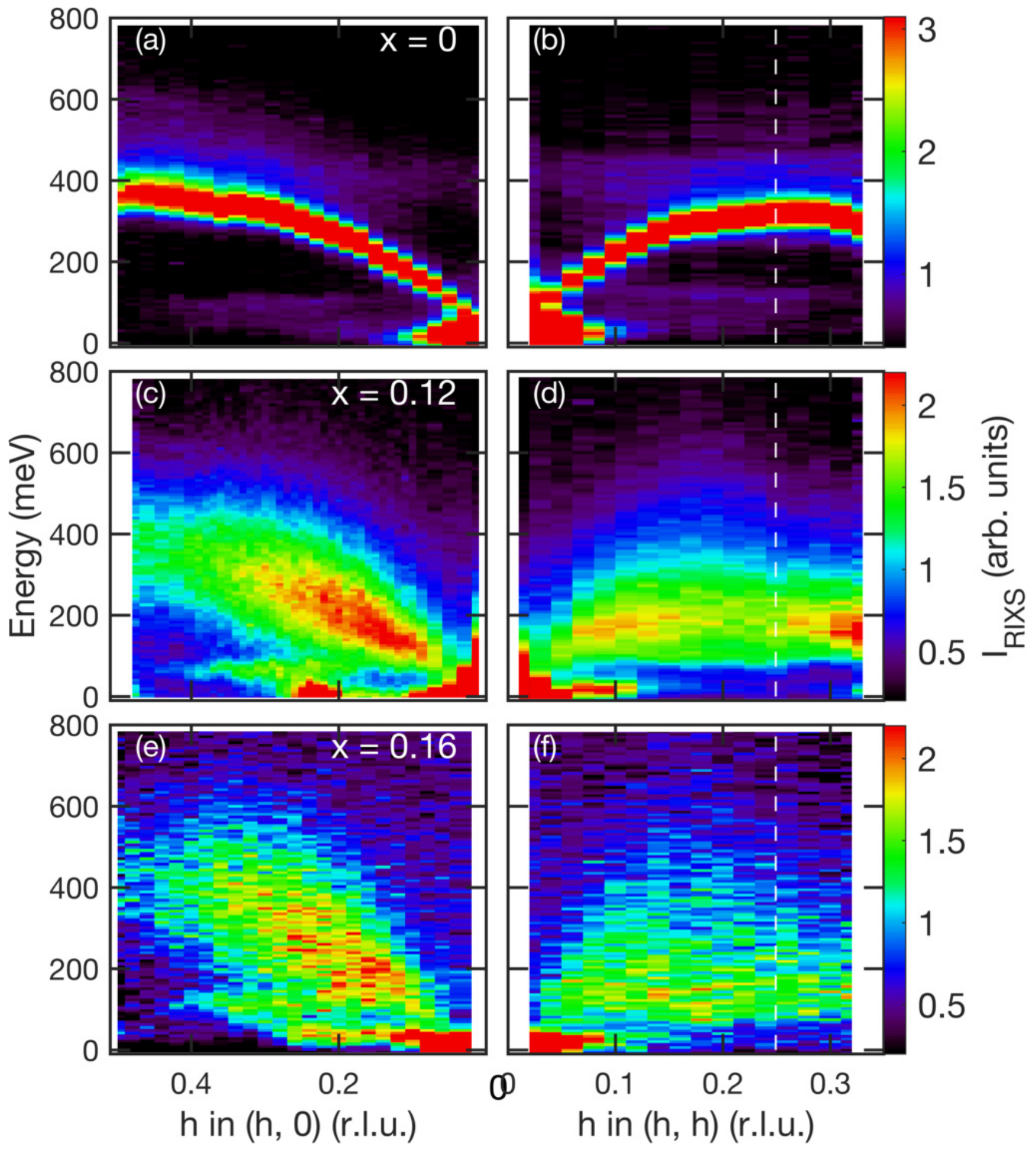}
\end{center}
\caption{RIXS spectrum of La$_{2-x}$Sr$_x$CuO$_4$ from Robarts {\it et al.\/}~\cite{Hayden19}. Note how the sharp spin wave excitation at $x=0$ turns into broad continua at non-zero $x$.
The latter have been argued by Bonetti {\it et al.\/}~\cite{BCS24} to be the spinons needed for the theory of the pseudogap metal. See also the comparison with numerics in Fig.~\ref{fig:Becca}.}
\label{fig:spinons}
\end{figure}

We can ask for experimental evidence for the spinons of the spin liquid underlying the pseudogap metal. 
We have argued recently \cite{BCS24} that such evidence is readily available in existing observations:
spinons are the most natural interpretation of RIXS measurements of the higher energy spin fluctuation spectrum in the cuprates \cite{Keimer11,Hayden19}.
As shown in Fig.~\ref{fig:spinons}, the doped cuprates have a broad continuum of high energy spin excitations, which become sharp spin waves in the zero doping limit. These continuum excitations have been labeled ``intense paramagnons'' \cite{Keimer11}, but this appears implausible because of the absence of a large Fermi surface in the underdoped regime. A spin wave interpretation is also not tenable in the absence of antiferromagnetic order.  We also note the recent NMR observations of a spin gap which have been associated with `short-range spin singlets' \cite{Julien25}.

Indeed, the cuprate observations are the analog of those in the triangular lattice antiferromagnet KYbSe$_2$ \cite{Scheie24}. As discussed in Section~\ref{sec:intro2} and Fig.~\ref{fig:KYbSe2}, KYbSe$_2$ displays a continuum at high energies which has been modeled in a theory of bosonic spinons \cite{Batista18,Scheie24}, along with weak antiferromagnetic order at low temperatures. 
\begin{figure}
\begin{center}
\includegraphics[width=3.8in]{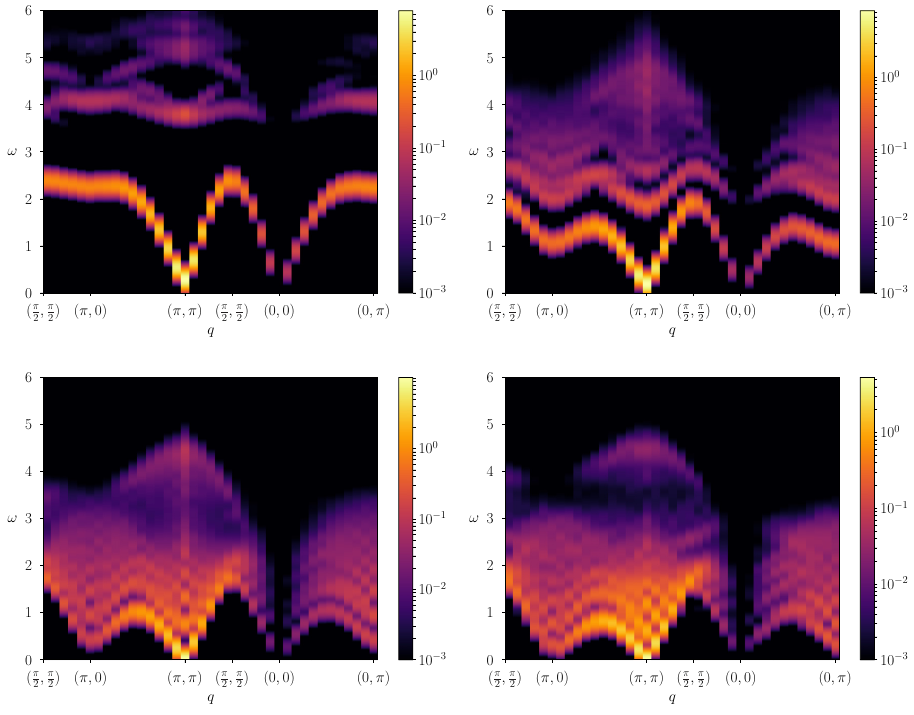}
\end{center}
\caption{Dynamic structure factor of the square lattice Heisenberg model with first ($J_1$) and second ($J_2$) neighbor exchange interactions from Ferrari and Becca \cite{Becca18},
$J_2/J_1 = 0$
(upper left), 0.3 (upper right), 0.45 (lower left), and 0.55 (lower right). Note the evolution from sharp spin-wave modes at $J_2/J_1=0$ to spinon continua at larger $J_2/J_1$. Compare to the evolution of the cuprate spectrum in Fig.~\ref{fig:spinons} with increasing doping between momentum $q = (0,0)$ and $q=(0,\pi)$ and between $q = (0,0)$ and $q=(\pi/2,\pi/2)$.}
\label{fig:Becca}
\end{figure}
Instead the observations of Robarts {\it et al.}~\cite{Hayden19} are naturally interpreted as spinons. Note the evolution of their observations with increasing doping from a sharp spin-wave spectrum in Figs.~\ref{fig:spinons}a,b to a broad continuum in Figs.~\ref{fig:spinons}c-f. This matches Fig.~\ref{fig:Becca}, which shows the evolution of the dynamic spin structure factor of the insulating square lattice antiferromagnet with increasing second neighbor exchange $J_2$, as computed by Ferrari and Becca~\cite{Becca18} using fermionic spinons (see also the connection of spinons in the unfrustrated square lattice to the quasi-one-dimensional limit \cite{Ronnow15,Ronnow25}). There is also a correspondence in the energy scale, measured in units of the first neighbor exchange $J_1 \approx 130$ meV \cite{Coldea01}.

We recall evidence via photoemission \cite{Johnson11} and ADMR \cite{Ramshaw20} for small pocket Fermi surfaces in the pseudogap regime of the hole-doped cuprates. 
Evidence for small Fermi surfaces at low doping has also emerged in quantum simulations using ultracold atoms \cite{Koepsell21,Chalopin24}, which show a clear transformation from a `polaronic metal' to a FL with increasing doping. Numerical computations on ancilla wavefunctions of FL* \cite{Iqbal24,HenryShiwei24} agree well with these observations. 

Finally, as we have shown recently \cite{LPS24} (see discussion below (\ref{bosonprop})), the transition from FL to FL* has a unique signature in having a singular enhancement of thermopower in a `skewed marginal Fermi liquid' \cite{Georges_Skewed} in the presence of spatial disorder, and this is consistent with observations in the hole-doped cuprates \cite{Collignon21,Taillefer_Seebeck_PRX_2022}. 

\subsection{Theory of the FL* phase}
\label{sec:fls}

\begin{figure}
\begin{center}
\includegraphics[width=4.4in]{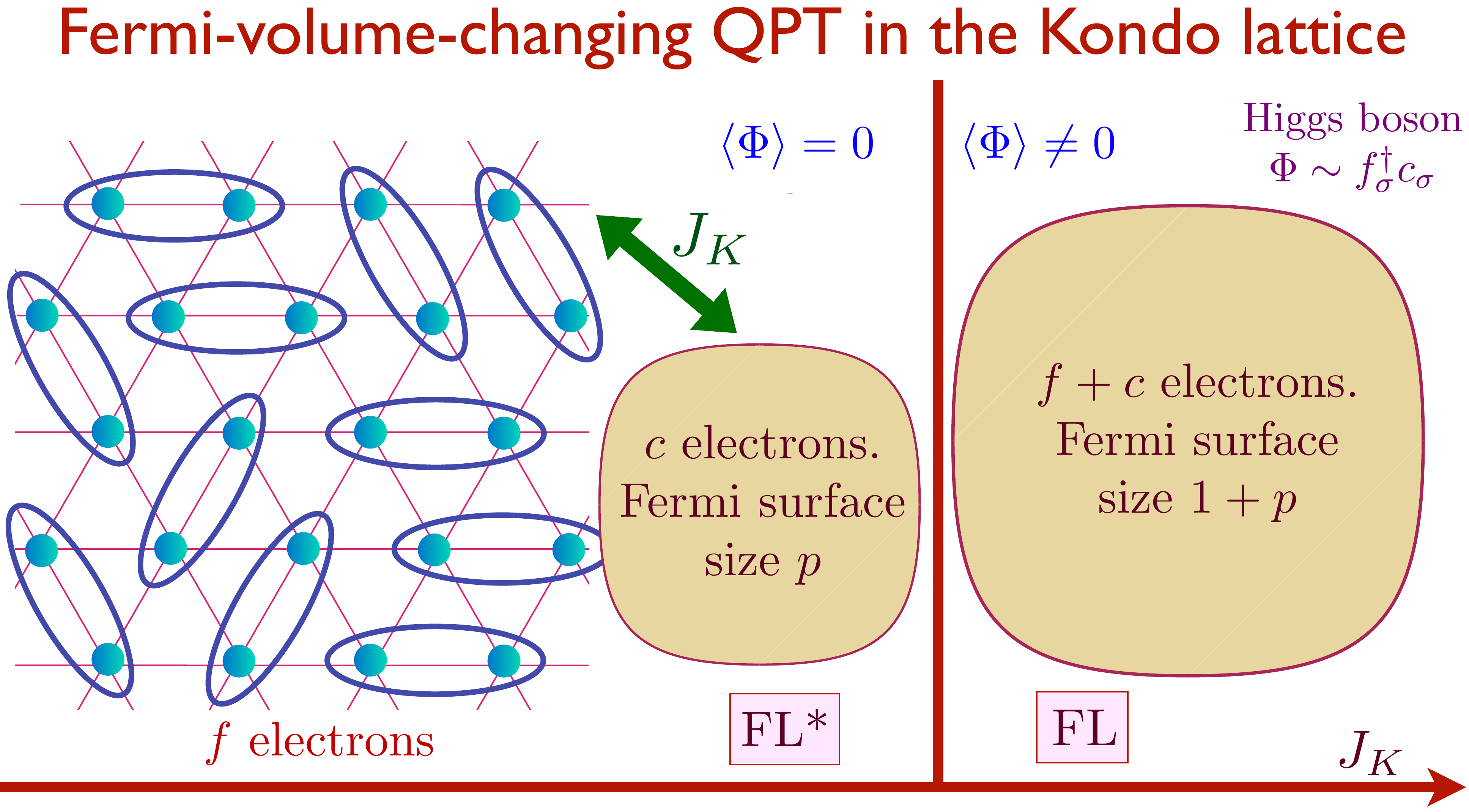}
\end{center}
\caption{Phases of a Kondo lattice model with $f$-electron spins coupled to a conduction band of $c$ electrons of density $p$. At large Kondo coupling, $J_K$, the condensation of a Higgs boson $\Phi$ leads to the conventional FL which obeys the Luttinger count. The FL* phase is obtained when $\Phi$ is uncondensed, and the $f$ electrons form a spin liquid whose gauge symmetry is preserved. Maksimovic {\it et al.\/} \cite{Analytis22} present evidence for such a transition in CeCoIn$_5$.}
\label{fig:kondo}
\end{figure}

It is a relatively simple matter to obtain a FL* phase in a Kondo lattice model, as illustrated in Fig.~\ref{fig:kondo}.
In the FL* phase, the $f$ electrons realize the spin liquid, while the $c$ conduction electrons form a Luttinger-volume-violating `small' Fermi surface. The transition from FL* to FL is driven by the condensation of a Higgs boson $\Phi \sim f^\dagger c$ carrying a fundamental charge of the emergent gauge field of the spin liquid. 

\begin{figure}
\begin{center}
\includegraphics[width=5in]{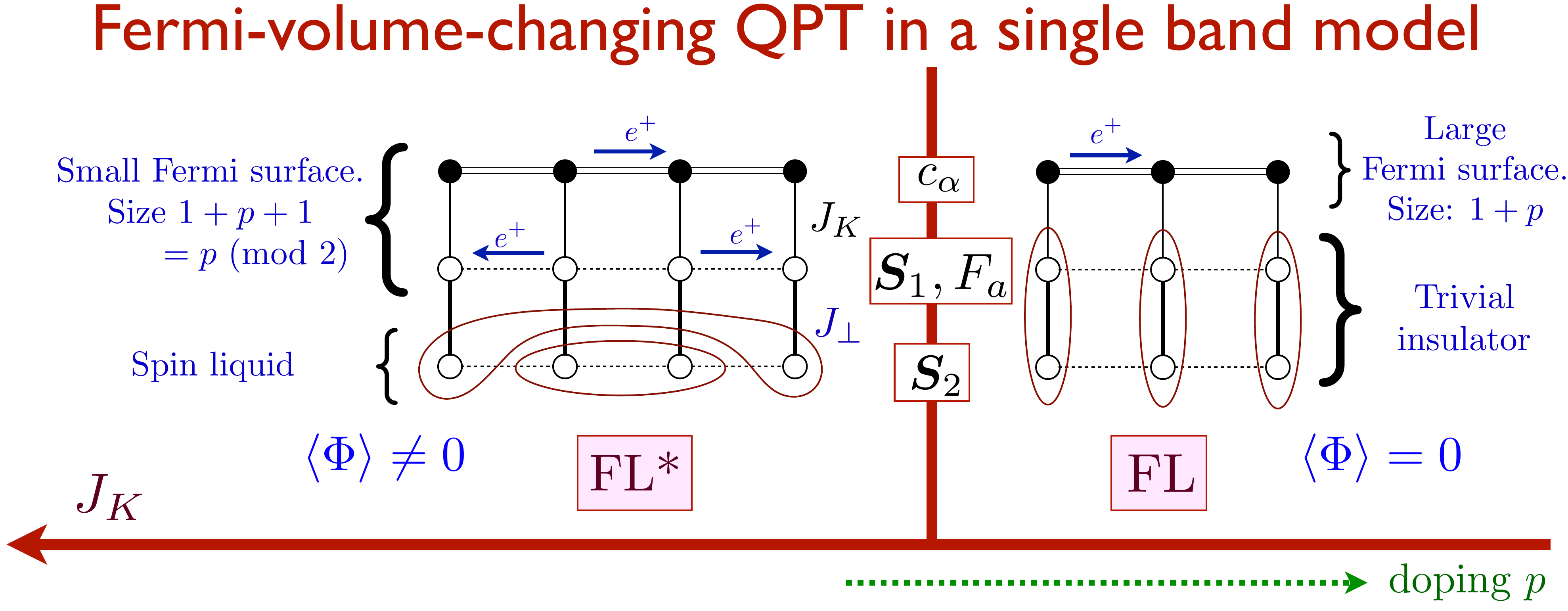}
\end{center}
\caption{Ancilla theory of a FL* phase in a single band model. After a canonical transformation, a Hubbard-like model of interacting electrons $c_\alpha$ is mapped to a model of free electrons $c_\alpha$ coupled to a bilayer antiferromagnet of spins ${\bm S}_1$ and ${\bm S}_2$. In the FL* phase, the ${\bm S}_2$ spins form a spin liquid, while
the ${\bm S}_1$ spins hybridize with $c_\alpha$ after the condensation of a Higgs boson $\Phi \sim F^\dagger c$ (where the ${\bm S}_1$ spins are represented by $F_a$ fermionic partons).}
\label{fig:ancilla}
\end{figure}

\begin{figure}
\begin{center}
\includegraphics[width=4.5in]{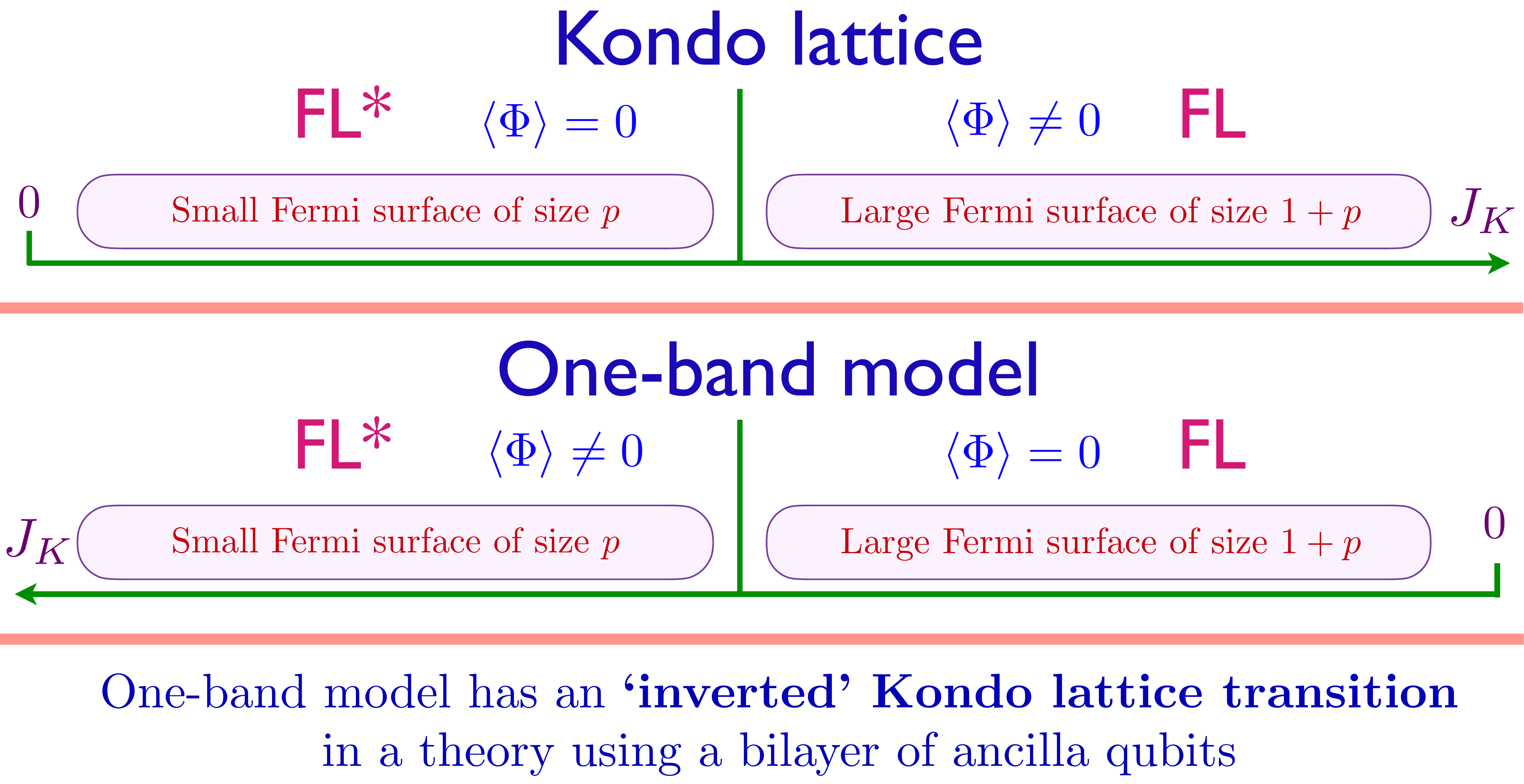}
\end{center}
\caption{Summary of FL* and FL phases of Kondo lattice and one-band models.}
\label{fig:inversion}
\end{figure}

Obtaining a FL* phase in a single band model is somewhat more subtle \cite{Qi10,Punk15,Henning24} (see Fig.~\ref{fig:holon}), and we illustrate the ancilla approach \cite{ZhangSachdev_ancilla,ZhangSachdev_ancilla2,YaHui24} in Fig.~\ref{fig:ancilla}.
Photoemission spectra obtained by this approach have been successfully compared with experiments in Ref.~\cite{Mascot22}.
For our purposes, it is sufficient to note that the FL to FL* transition is also driven by the condensation of a Higgs boson $\Phi$, but the transition is now `inverted' with respect to the Kondo lattice model, as shown in Fig.~\ref{fig:inversion}.

\subsection{FL-FL* QPT}

We now address the nature of the FL-FL* QPT in Fig.~\ref{fig:qpts}, in the presence of spatial disorder. We will do this for the case of the Kondo lattice transition in Fig.~\ref{fig:kondo} only, as the results for the single-band model of Fig.~\ref{fig:ancilla} differ only in their gauge structure \cite{SSORE,LPS24,ZhangSachdev_ancilla,ZhangSachdev_ancilla2,Zou20}, and this is unimportant at the level of the SYK-type equations in (\ref{SYKeq}). The theory is similar in structure to that for the symmetry breaking case in (\ref{e11}), with the Higgs field $\Phi$ replacing the SDW order parameter $\phi$: we have the imaginary time Lagrangian
for a 2dYSYK theory:
\begin{align}
 \mathcal{L}_2 & = \sum_{\bm k} c_{{\bm k}\sigma}^\dagger \left( \frac{\partial}{\partial \tau} + \varepsilon ({\bm k})\right) c_{{\bm k}\sigma} + \sum_{\bm k} f_{{\bm k}\sigma}^\dagger \left( \frac{\partial}{\partial \tau} + \varepsilon_1 ({\bm k})\right) f_{{\bm k}\sigma} \nonumber \\
& + \int d^2 {\bm r} \Bigl\{ s \,|\Phi({\bm r})|^2 +  [g + {\color{purple} g'({\bm r})}] \, c_{\sigma}^{\dagger}({\bm r})  f_{\sigma} ({\bm r}) \, \Phi({\bm r}) + \mbox{H.c.} \nonumber \\
& ~~~~~~~~~~+ K \, | \nabla_{\bm r} \Phi ({\bm r})|^2 +u  \,|\Phi({\bm r})|^4  + {\color{purple} v({\bm r})} c_\sigma^\dagger ({\bm r}) c_\sigma  ({\bm r}) \Bigr\}\,.
\label{e22}
\end{align}
where we have omitted the emergent gauge field under which the Higgs field $\Phi$ and the spinons $f$ are charged \cite{TSSSMV04}. It is remarkable that, in the presence of spatial disorder, the critical properties of the QPT without symmetry breaking described by $\mathcal{L}_2$ are essentially the same as those of the QPT with symmetry-breaking described by $\mathcal{L}_1$ in (\ref{e11}):
hence the claim of a {\it universal\/} theory of strange metals in Ref.~\cite{Patel:2022gdh}. There is, however, one important difference between $\mathcal{L}_1$ and $\mathcal{L}_2$. The Higgs field $\Phi$ is complex, and there is a pronounced particle-hole asymmetry in $\mathcal{L}_2$, which is absent for $\mathcal{L}_1$ \cite{LPS24}. Specifically, in the regime of extended boson eigenstates (see Fig.~\ref{fig:eigenmodes}), the renormalized propagator for the Higgs boson has the form \cite{Aldape22,LPS24}
\begin{align}
D(i\Omega, {\bm q}) \sim \frac{1}{K {\bm q}^2 + c_1 |\Omega| - i c_2  \Omega + m^2 (T)} \label{bosonprop}
\end{align}
Here $c_1$ term is the usual Landau damping in a disordered metal. The crucial particle-hole asymmetry is induced by $c_2$: this vanishes for ${\mathcal{L}}_1$ by time-reversal or inversion symmetry, but is non-zero for $\mathcal{L}_2$. Consequently, there are singular `skewed marginal Fermi liquid' contributions to thermopower for the FL-FL* QPT \cite{LPS24}, as we noted above.

\begin{figure}
\begin{center}
\includegraphics[width=4.5in]{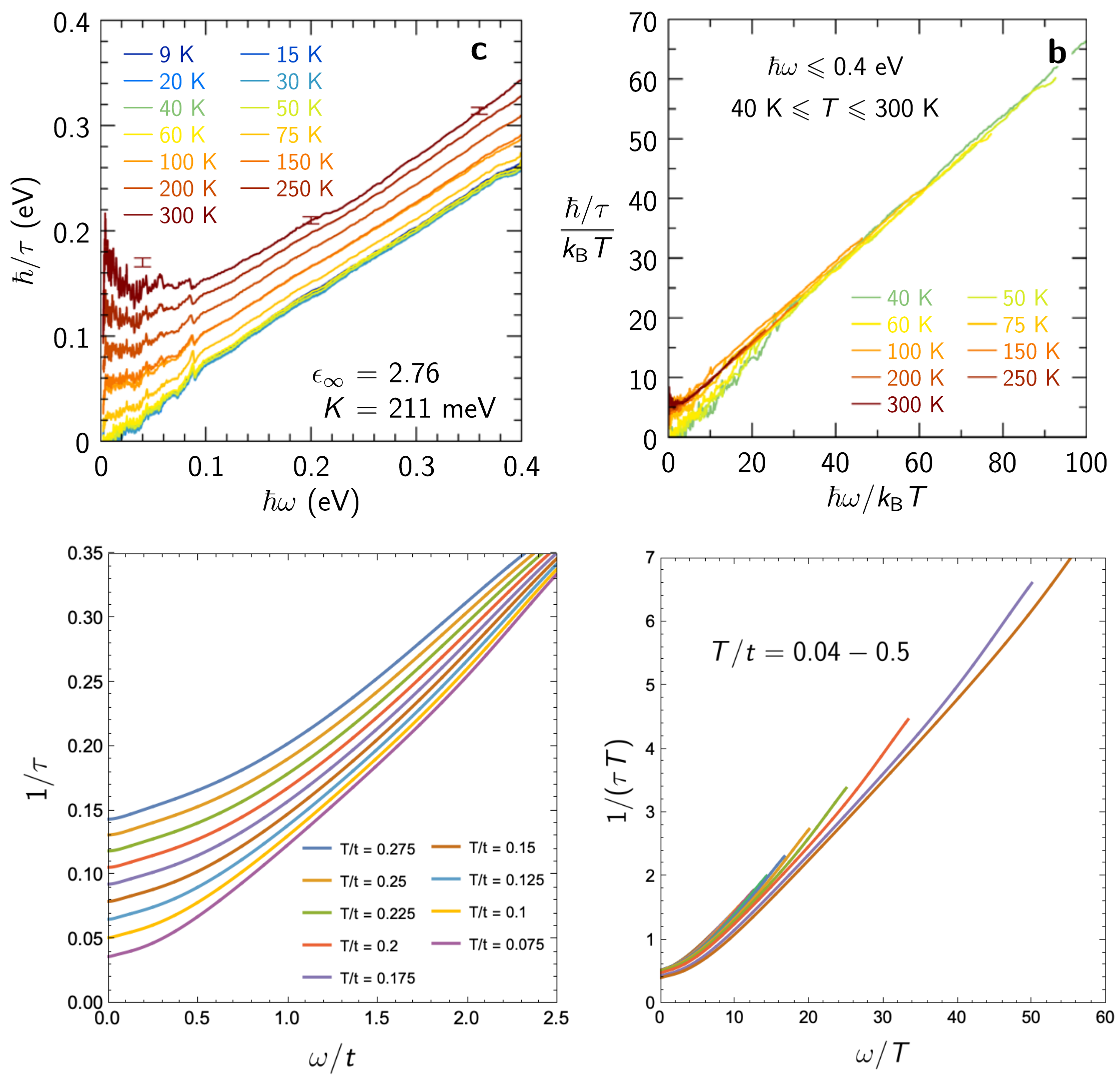}
\end{center}
\caption{The top panels display the measurements of transport relaxation time, $\tau$, obtained from the optical conductivity in Michon {\it et al.\/}~\cite{Michon22}.
The bottom panels show computation of the same quantity in the 2dYSYK model by Li {\it et al.\/}~\cite{Li:2024kxr}.
}
\label{fig:tau}
\end{figure}
We are interested in applying $\mathcal{L}_2$ to the `fan' at higher temperatures, and so we can focus on the case where all the fermionic and bosonic eigenmodes are extended, and we apply self-averaging SYK techniques. After adding a large number of flavors and making the couplings random in flavor space, as in the SYK model, we can obtain  the fully self-consistent solution SYK-type equations \cite{Esterlis21,Patel:2022gdh}:
\begin{align}
&\Sigma(\tau,{\bm r}) = g^2 D(\tau,{\bm r})G(\tau,{\bm r}) + {\color{purple} v^2 \delta^2({\bm r})} G(\tau,{\bm r})  + {\color{purple} {g'}^2 \delta^2({\bm r})}G(\tau,{\bm r })D(\tau,{\bm r}), \nonumber \\
& \Pi(\tau,{\bm r}) = -g^2 G(-\tau,-{\bm r})G(\tau,{\bm r}) - {\color{purple} {g'}^2G(-\tau,{\bm r}) \delta^2({\bm r})} G(\tau,{\bm r}), \nonumber \\
&G(i\omega,{\bm k}) = \frac{1}{i\omega-\varepsilon({\bm k})-\Sigma(i\omega,{\bm k})}, \nonumber \\
&D(i\Omega,{\bm q}) = \frac{1}{\Omega^2/c^2+K{\bm q}^2+s-\Pi(i\Omega,{\bm q})}\,,
\label{SYKeq}
\end{align}
where $G$ is the electron Green's function, $D$ is the boson Green's function, and $\Sigma$ and $\Pi$ are the corresponding self energies. The equations have been written down for the simpler case where $\varepsilon ({\bm k}) = \varepsilon_1 ({\bm k})$, which applies to the SDW case in (\ref{e11}); in the presence of elastic scattering from {\color{purple} $v({\bm r})$}, the solutions are insensitive to the precise Fermi surfaces.
The solution of (\ref{e22}) by (\ref{SYKeq}) should be valid as long as the $\Phi$ eigenmodes remain extended. The equations (\ref{SYKeq}) have been solved at $g=0$ in Refs.~\cite{Aldape22,Esterlis21,Patel:2022gdh,Li:2024kxr,LPS24}.
The solutions
display marginal Fermi liquid behavior in the fermion Green's function, and linear-in-$T$ resistivity. Importantly, the Planckian behavior of the relaxation time of the SYK model in (\ref{e1}) is preserved (see Fig.~\ref{fig:tau}); but the zero temperature entropy of the SYK model in (\ref{e2}) is not. Instead, the 2dYSYK model solution via (\ref{SYKeq})  displays a $\sim T \ln (1/T)$ entropy at the critical point, again consistent with observations. 

\section{Confinement crossover}
\label{sec:confinement}

Finally, we turn to the confinement crossover in Fig.~\ref{fig:qpts}, from the pseudogap metal state at higher temperature, to the symmetry-broken states without fractionalization at low temperatures. 

We need to begin by discussing the nature of the pseudogap metal itself, and in particular the nature of the underlying spin liquid. An important subtlety is that the square lattice spin liquid has dual representations employing fermionic or bosonic spinons.

Christos {\it et al.\/} \cite{Christos23} have proposed the $\pi$-flux spin liquid with Dirac fermion spinons \cite{Affleck1988}.
The fermionic spinons $f_{i \sigma}$ are obtained by transforming the electrons $c_{i \sigma}$ to a rotating reference frame 
in pseudospin space \cite{Scheurer18}
\begin{align}
\left(
\begin{array}{cc}
c_{i \uparrow} & - c_{i \downarrow}^\dagger \\
c_{i \downarrow} & c_{i \uparrow}^\dagger
\end{array}
\right) \sim \left(
\begin{array}{cc}
f_{i \uparrow} & - f_{i \downarrow}^\dagger \\
f_{i \downarrow} & f_{i \uparrow}^\dagger
\end{array}
\right)
\left(
\begin{array}{cc}
B_{i 1}^\ast & B_{i 2} \\
-B_{i 2}^\ast & B_{i 1}
\end{array}
\right)\,, \label{frac1}
\end{align}
where $(B_{i1}, B_{i2})$ are charge $e$ spinless bosons. This description has an emergent SU(2) gauge field, and $B$ is an SU(2) fundamental Higgs field \cite{Li_06}. This approach has been developed for a theory of the pseudogap as a FL* with a $\pi$-flux spin liquid, and its lower $T$ confinement transitions in Refs.~\cite{Christos23,Christos24,Luo24}.

\begin{figure}
\begin{center}
\includegraphics[width=4.5in]{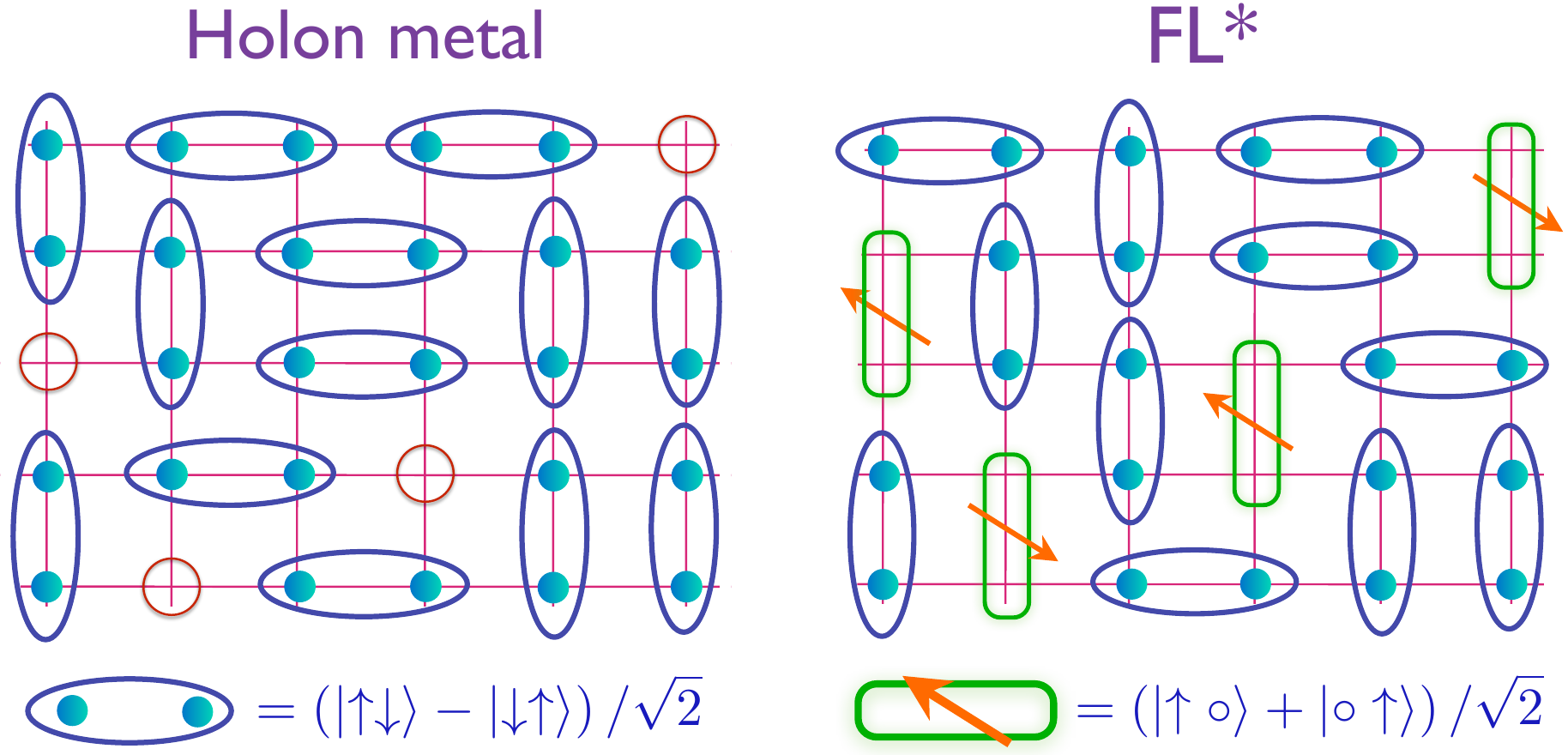}
\end{center}
\caption{Cartoon representations of the holon metal and FL* states. The holon metal has spinless charge $+e$ fermions, while FL* has spin-1/2 charge $+e$ fermions, above the same underlying spin liquid with short-range singlet valence bonds. The FL* is shown as a snapshot of the quantum dimer model of Punk {\it et al.\/} \cite{Punk15}.}
\label{fig:holon}
\end{figure}

Using fermion-boson duality, the $\pi$-flux spin liquid was argued by Wang {\it et al.\/} \cite{DQCP3} to have dual description as the $\mathbb{CP}^1$ theory of bosonic spinons \cite{NRSS89,NRSS90}. 
This bosonic spinon description is obtained by the complementary representation of fractionalization by a rotating reference frame in spin space \cite{SDW09,Topo_PNAS,SS_ROPP19,Bonetti22}
\begin{align}
\left(
\begin{array}{cc}
c_{i \uparrow} & - c_{i \downarrow}^\dagger \\
c_{i \downarrow} & c_{i \uparrow}^\dagger
\end{array}
\right) \sim 
\left(
\begin{array}{cc}
z_{i \uparrow} & - z_{i \downarrow}^\ast \\
z_{i \downarrow} & z_{i \uparrow}^\ast
\end{array}
\right) 
\left(
\begin{array}{cc}
\psi_{i +} & - \psi_{i -}^\dagger \\
\psi_{i - } & \psi_{i +}^\dagger
\end{array}
\right)\,, \label{frac2}
\end{align}
where $z_{i \sigma}$ are the bosonic spinons, and $\psi_{\pm}$ are charge $-e$ spinless fermions. 
This description has a different emergent SU(2) gauge field, but it is higgsed down to U(1) to yield the $\mathbb{CP}^1$ theory for the $z_\sigma$. The $\psi_p^\dagger$ are the holons in a holon metal carrying charges $p = \pm 1$ under the emergent U(1) \cite{ACL08,SDW09,Topo_PNAS,SS_ROPP19, Bonetti22}; see Fig.~\ref{fig:holon}. The possible low $T$ fates of the holon metal have been discussed in Ref.~\cite{SSST19}.

Numerical evidence for these dual descriptions appears in recent fuzzy sphere results \cite{FuzzySO5}, and it was argued \cite{BCS24} from these results that bosonic spinons provided a better description of the dynamic spin structure factor of this spin liquid.
Which of the fermionic spinon/bosonic chargon in (\ref{frac1}) or the bosonic spinon/fermionic chargon in (\ref{frac2}) descriptions is more appropriate for the electronic spectrum of the thermally fluctuating pseudogap metal has not been resolved at present; indeed, both could be acceptable. Irrespective of an eventual resolution, we have argued in Section~\ref{sec:qptwsb} that an underlying FL-FL* QPT provides at least a satisfactory description of the quantum critical fan \cite{LPS24}.

\begin{figure}
\begin{center}
\includegraphics[width=4.5in]{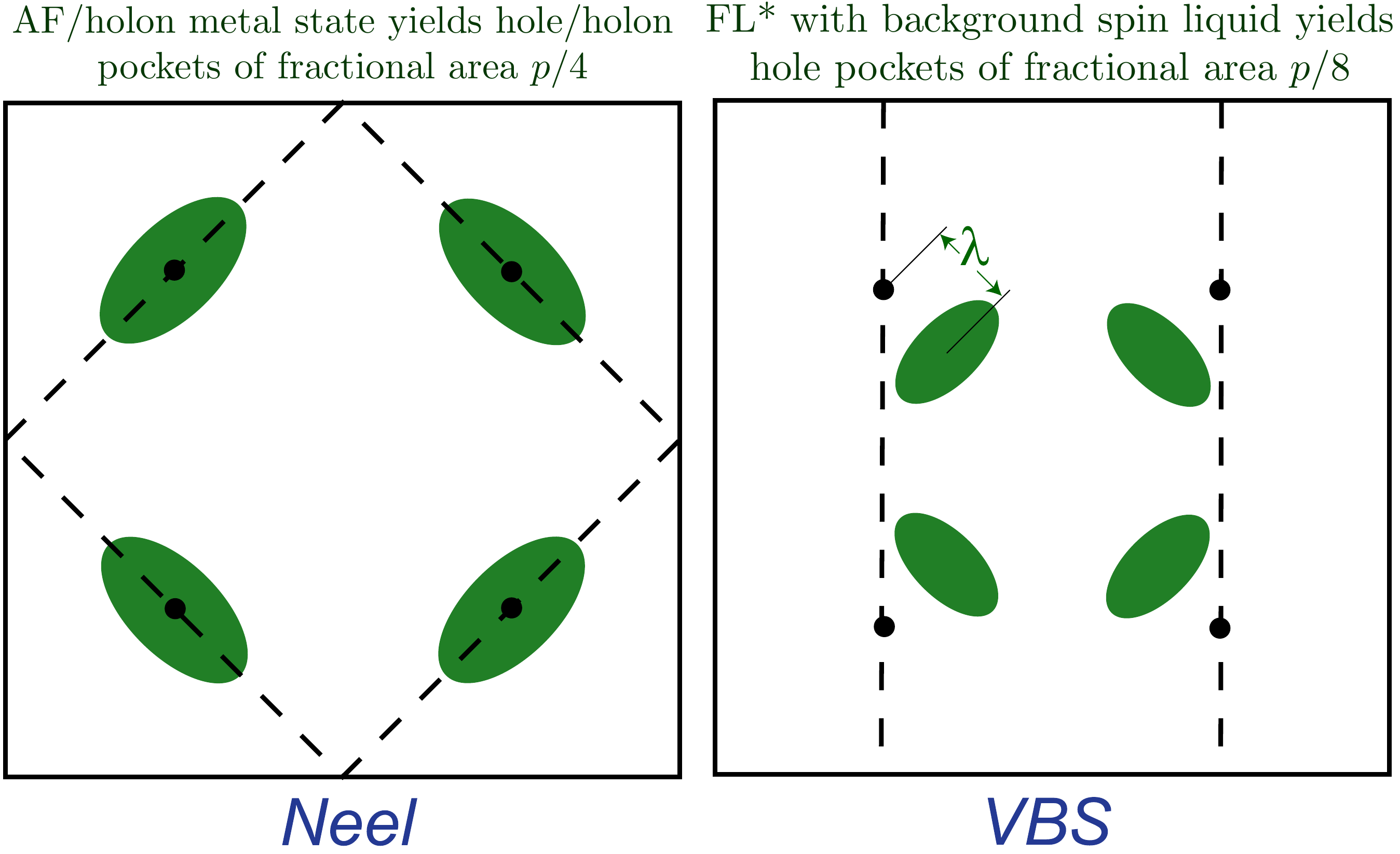}
\end{center}
\caption{From Kaul {\it et al.} \cite{Kaul07}, with added text on top. Fermi surfaces in the square lattice Brillouin zone obtained by doping an insulating antiferromagnet in the vicinity of a quantum phase transition from the N\'eel state to the valence bond solid (VBS). 
The N\'eel state has 2 pockets of spinful quasiparticles in the magnetic Brillouin zone, and so each hole pocket each occupies a fractional area $p/4$. (The holon metal has 4 pockets of spinless quasiparticles, and so the fractional area of each pocket remains $p/4$).
The VBS order appears at long distances in the insulator above an underlying $\pi$-flux spin liquid \cite{NRSS89,NRSS90,DQCP3}, and the ultimate broken translational symmetry plays no role in Fermi surface structure in the computation in Kaul {\it et al.} \cite{Kaul07}. Consequently the right panel applies also to a FL* metal: there are 4 pockets of spinful quasiparticles, and so the fractional area of each pocket is $p/8$, as stated by Kaul {\it et al.\/} \cite{Kaul07}, and close to the value observed by Chan {\it et al.\/} \cite{Yamaji24}.}
\label{fig:kaul}
\end{figure}

It is worth mentioning here an important low $T$ difference between FL* and the holon metal which could be significant for transport experiments. As the holons are spinless, the area enclosed by a holon pocket is twice that of the hole pocket in FL* \cite{Kaul07}, and the holon pocket area is the same as that of the hole pocket in the ordered antiferromagnet (see Fig.~\ref{fig:kaul}). 
Very recent observations of the Yamaji effect \cite{Yamaji24} in the single layer cuprate HgBa$_2$CuO$_{4+\delta}$ at doping $p=0.1$ support the FL* theory of the pseudogap: there are 4 hole pockets of spin-1/2 fermions in FL*, and so each hole pocket occupies a fraction $p/8 = 0.0125$ of the Brillouin zone, to be compared to the observed value $\approx 1.3$\% \cite{Yamaji24}.

In addressing the onset of confinement at low $T$, we can exploit the duality of Wang {\it et al.\/} \cite{DQCP3} to choose the formulation best suited for the confinement being considered.
In the fermionic spinon description, the Higgs boson $B$ is uncondensed in the pseudogap metal, and there is a relatively straightforward method to understand confinement  
via the Higgs condensation of $B$.
A number of puzzles on the cuprate phase diagram can be addressed by this method:
\begin{itemize}
\item For a suitable Higgs potential \cite{Christos23}, the condensation of $B$ leads to $d$-wave superconductivity, with four nodal fermionic quasiparticles. These quasiparticles have anisotropic velocities, similar to those in the BCS $d$-wave superconductor \cite{Chatterjee16,Christos24}.
\item When applied to the $d$-wave superconductor in the electron-doped cuprates, we obtain \cite{Christos24} a non-monotonic momentum space evolution of the gap away from the anti-nodal point, as is observed \cite{Shen_2023}. This theory also makes the remarkable prediction that 4 nodal quasiparticles will emerge in the $d$-wave superconductor even when the Brillouin zone diagonals of the parent pseudogap metal are gapped \cite{Christos24}.
\item For a suitable Higgs potential \cite{Christos23}, the condensation of $B$ can lead to charge order. This has successfully described the evolution of the electronic spectrum from the pseudogap metal to the single electron pocket observed in high magnetic field quantum oscillation experiments \cite{BCS24}. It could also be used to model the modulations observed in scanning tunneling microscopy experiments \cite{Davis07,Davis12,Davis16,Davis20}.
\item A long-standing puzzle in the cuprates has been the nature of vortices in the underdoped $d$-wave superconductor. The Bogoliubov-de Gennes theory, as applied by Wang and MacDonald \cite{WM95}, predicts a large zero bias peak in the electronic local density of states (LDOS) at the vortex center. The Wang-MacDonald peak has finally been observed \cite{Renner21}, but only in heavily overdoped Bi$_{2}$Sr$_{2}$CaCu$_{2}$O$_{8 +\delta}$. No such peak is seen in the underdoped cuprates: instead, the pioneering scanning tunneling microscopy observations of Hoffman {\it et al.} \cite{Hoffman02} observed sub-gap peaks at $\pm$6-9 meV in the LDOS which exhibit periodic spatial modulations in a `halo' around the vortex core. A similar structure has been obtained in recent computations with the Higgs field $B$ \cite{ZhangVortex24}, along with several other features of the vortex spectrum, but in a simplified model of the Higgs potential leading to spatial modulations of period 2 lattice spacings.
\end{itemize}
The confinement crossover from the pseudogap metal to the stripy SDW state at low temperatures remains to be understood \cite{Joshi23}. In the fermionic spinon approach, it involves a delicate interplay between the charge order associated with the condensation of $B$, and the confinement of the fermionic spinons via the SU(2) gauge field \cite{Luo24}. The bosonic spinon approach \cite{Topo_PNAS,SS_ROPP19,Bonetti22,SSST19} is an alternative route to understanding this crossover, and
insights from numerical studies \cite{Georges_pseudogap_24} should also be helpful.

\section*{Acknowledgments}
I thank Pietro Bonetti, Maine Christos, Ilya Esterlis, Haoyu Guo, Darshan Joshi, Chenyuan Li, Zhu-Xi Luo, Peter Lunts, Alex Nikolaenko, Jacopo Radaelli, Mathias Scheurer, J\"org Schmalian, Henry Shackleton, Maria Tikhanovskaya, Davide Valentinis, Jia-Xin Zhang, Ya-Hui Zhang, and especially Stephen Hayden  and Aavishkar Patel for recent collaborations on which 
this perspective is based.  I remain indebted to Jan Zaanen for freely sharing his creative insights over several decades of wide-ranging discussions on the cuprates, and many other aspects of physics. 
This research was supported by the U.S. National Science Foundation grant No. DMR-2245246, and by the Simons Collaboration on Ultra-Quantum Matter which is a grant from the Simons Foundation (651440, S. S.). 


\bibliography{zaanen}

\end{document}